\documentclass[referee]{raa}
\usepackage{graphicx,times}
\usepackage{natbib}
\usepackage{amssymb,amsmath}
\bibpunct{(}{)}{;}{a}{}{,}
\usepackage[scheme=plain]{ctex}
\usepackage[pagebackref=true]{hyperref}
\newcommand{\rev}[1]{{#1}}
\begin{document}

  \title{Multiple populations detection with the Chinese Space Station Survey Telescope main survey camera
}

   \volnopage{Vol.0 (202x) No.0, 000--000}      
   \setcounter{page}{1}          

   \author{Zhuohang Li\inst{1,2}, Xia Li\inst{1,2}, Hao Tian\inst{3,4}, Xin Zhang\inst{3}, Antonino P. Milone\inst{5,6}, Long Wang\inst{1,2}, Baitian Tang\inst{1,2}, Edoardo P. Lagioia \inst{7}
   \and Chengyuan Li\inst{1,2}\fnmsep\thanks{Shows the usage of elements in the author field}}

   \institute{School of Physics and Astronomy, Sun Yat-sen University, Daxue Road, Zhuhai, 519082, P.R. China         \\\email{lichengy5@mail.sysu.edu.cn}
    \and CSST Science Center for the Guangdong-Hong Kong-Macau Greater Bay Area, Zhuhai, 519082, China 
    \and National Astronomical Observatories, Chinese Academy of Sciences, Beijing 100101, China 
    \and Institute for Frontiers in Astronomy and Astrophysics, Beijing Normal University, Beijing, 102206, China
    \and Osservatorio Astronomico di Padova, Vicolo dell' Osservatorio 5, 35122 Padova, Italy
    \and Dipartimento di Fisica e Astronomia ``Galileo Galilei'', Univ. di Padova, Padova, Italy
    \and South-Western Institute for Astronomy Research, Yunnan University, Kunming 650500, P.R. China}
    
   \institute{School of Physics and Astronomy, Sun Yat-sen University, Daxue Road, Zhuhai, 519082, P.R. China; {\it lichengy5@mail.sysu.edu.cn}\\
    \and
        CSST Science Center for the Guangdong-Hong Kong-Macau Greater Bay Area, Zhuhai, 519082, China\\
    \and
        National Astronomical Observatories, Chinese Academy of Sciences, Beijing 100101, China \\
    \and
        Institute for Frontiers in Astronomy and Astrophysics, Beijing Normal University, Beijing, 102206, China\\
    \and
        Osservatorio Astronomico di Padova, Vicolo dell' Osservatorio 5, 35122 Padova, Italy\\
    \and
        Dipartimento di Fisica e Astronomia ``Galileo Galilei'', Univ. di Padova, Padova, Italy\\
    \and
        South-Western Institute for Astronomy Research, Yunnan University, Kunming 650500, P.R. China\\
\vs\no
   {\small Received 202x month day; accepted 202x month day}}

\abstract{The phenomenon of multiple stellar populations (MPs), characterised by star-to-star variations in light-element abundances, is a ubiquitous feature of globular clusters (GCs). While spectroscopic surveys have directly revealed these abundance anomalies, photometric studies, particularly with the \textit{Hubble Space Telescope} (HST), have been instrumental in characterizing MP sequences across the colour-magnitude diagrams (CMDs). However, the narrow field of view of the HST has restricted these studies to only a small portion of the star clusters, leaving the vast majority of the clusters' space unexplored. The upcoming \textit{Chinese Space Station Survey Telescope} (CSST), with its wide field of view and both UV and optical capabilities, will provide new opportunities for systematic MP studies. We aim to quantify the capability of the CSST wide-field camera in detecting and characterizing MPs in GCs through virtual simulations that closely mimic real observations. We performed comprehensive simulations using synthetic stellar population models incorporating different helium abundances ($\Delta{Y}$) and carbon-nitrogen-oxygen (CNO) variations. We simulated CSST observations for GCs at distances of about \rev{9.6 kpc and 20 kpc} with different exposure times. We evaluated the detection efficiency of MPs based on the CMDs constructed from the seven different bands of the CSST, as well as the pseudo colour-colour diagrams (chromosome diagrams) designed with the UV-optical combination. For a GC at a distance of approximately 9.6 kpc, the $NUV-u$ colour index of the CSST is highly sensitive to stellar populations with different $\Delta{Y}$ and CNO abundances in the CMD, providing colour separations of $\Delta$($NUV-u$) $\approx$ 0.16 (for red giants) to 0.44 mag (for dwarfs). We find that when the total exposure time in the UV band exceeds $\sim$1000 s and the exposure time in the optical band exceeds $\sim$300 s, the main survey camera of CSST is sufficient to resolve these MPs. \rev{Furthermore, extending the simulations to 20~kpc, a distance encompassing $\sim$80\% of Galactic GCs, we demonstrate that the SCam retains significant diagnostic power, resolving stellar populations with $\Delta Y \geq 0.06$~dex and $\delta[\mathrm{N/Fe}] \geq 0.64$~dex, and distinguishing MPs in the CMD down to $i \sim 19.5$~mag for clusters with substantial chemical dispersions.} The combination of NUV--u--g filters provides diagnostic capabilities comparable to HST's F275W--F336W--F438W system.CSST will be a powerful facility for MP studies, capable of efficiently surveying the entire spatial extent of several hundred star clusters in the local group. Its wide field of view (FoV) and multi-band capabilities will enable the first homogeneous MP census spanning the entire Milky Way and its neighbouring galaxies, significantly advancing our understanding of star clusters' formation and chemical evolution.
}

   \authorrunning{Li et al. }            
   \titlerunning{MPs detection with the CSST MS}  

   \maketitle

%
\section{Introduction}           
\label{sect:intro}
Globular Clusters (GCs), which are among the oldest stellar systems in the Milky Way, preserve numerous traces of the Galaxy's early evolution. For many years, stars in GCs were regarded as prototypes of simple stellar populations (SSP). They formed in a single star formation event, with all stars having similar ages and initial chemical compositions within a narrow range. This view was based on fundamental stellar evolution theory: stellar winds, radiation, and supernovae from rapidly evolving massive stars would disperse the gravitationally bound molecular cloud, halting subsequent star formation by removing the gas reservoir.

Chemical anomalies in globular clusters were first identified in the 1970s and possibly earlier \citep[e.g.,][]{1973ApJ...186..725O, 1981ApJ...248..177N}. Since the 1980s, a series of spectroscopic observations from ground-based telescopes have confirmed significant star-to-star chemical abundance variations (in particular the C, N, O, Na abundances) in GC stars, marking the beginning of the era of multiple population (MP) studies in star clusters \citep{1987ApJ...313L..65N, 1989PASP..101.1083S, 1991ApJ...373..482B}. This phenomenon appears in almost all evolutionary stages, including the main sequence (MS), subgiant branch (SGB), red giant branch (RGB), horizontal branch (HB) and even asymptotic giant branch (AGB). This indicates that the MPs are independent of stellar evolution \citep{1998MNRAS.298..601C, 2001A&A...369...87G, 2005AJ....130.1177C,2021csss.confE.137L}. However, measuring chemical abundances in GCs is challenging. It is difficult to remove stray light from nearby stars in crowded fields to obtain clean spectra. Therefore, most early studies focused on the outer regions where stellar density is lower. Since many theoretical models predict that second-generation stars form in the dense cluster centre. Thus, the lack of samples in the core region hinders the development of these models \citep[e.g.,][]{2007A&A...464.1029D, 2008MNRAS.391..825D}. 

To eliminate these difficulties, researchers turned to point-spread-function based photometry to study MPs. The vast majority of breakthroughs in this field originate from observations with the \textit{Hubble Space Telescope} (HST). The underlying rationale is that variations in light elements such as C, N, and O are detectable in filters blueward of $\sim$$4000$~\AA: OH absorptions dominates near $\lambda \approx 3000$~\AA, while NH, CN, and CH absorptions are centred at $\lambda = 3370$~\AA, $3883$~\AA, and $4300$~\AA, respectively. Moreover, the Mg~\textsc{ii} doublet at $2795/2805$~\AA is one of the most important UV absorption features, detectable with appropriate UV filters. Photometric comparisons using these bands can distinguish chemical subpopulations. Furthermore, given that helium is the second most abundant element in stars, variations in its mass fraction have a direct impact on the entire stellar interior structure. Helium enrichment reduces atmospheric opacity and increases the mean molecular weight, making He-rich stars appear bluer and brighter. These structural modifications manifest as distinct observational signatures, such as the splitting of the main sequence, which can be directly detected via high-precision photometry \citep[e.g.,][]{2007ApJ...661L..53P, 2012ApJ...744...58M}. As \cite{2017MNRAS.464.3636M} shows, 'chromosome maps' (ChMs) constructed from photometric indices clearly separate 1P and 2P stars, underscoring the power of UV--optical telescopes in probing the MPs phenomenon.

The HST has greatly advanced our understanding of MPs in at least three aspects. First, its photometry clearly revealed split MSs. \rev{Such splits} are mainly caused by variations in helium abundance ($\Delta{Y}$) \citep{2007ApJ...661L..53P}. This is of \rev{particularly} valuable because direct helium measurement is difficult with spectroscopy as most stars in GCs are cool, while helium lines require high temperatures. Second, HST data have demonstrated the evolutionary continuity of MPs, showing that distinct populations form continuous sequences extending from the MS to the RGB.\citep{2017MNRAS.464.3636M} Finally, high-precision astrometry allowed us to study the spatial and kinematic differences of these populations. Photometric and spectroscopic \rev{studies} revealed that such properties vary significantly between clusters \citep{2023MNRAS.520.1456L}. 

Thanks to photometric and spectroscopic surveys, it is recognised that the MP phenomenon is unique to GCs (or at least old, massive clusters). The most typical characteristic of second-generation stars is their significant abundance anomalies in He, C, N, and O compared to normal stars of the same metallicity. In some clusters, these anomalies extend to heavier elements, manifesting as the well-known Mg-Al anticorrelation \citep[e.g.,][]{2009A&A...505..117C}. Such chemically anomalous stars are rarely found in open clusters or the Galactic field \citep{2012A&ARv..20...50G, 2012A&A...548A.122B, 2011A&A...534A.136M, 2020MNRAS.491..515M}. Although abundance patterns vary from cluster to cluster, all GCs exhibit the same basic pattern: stars can be divided into a first stellar population (1P) enriched in C and O, and a second population (2P ) enriched in helium, N, and sodium (Na) but depleted in C and O. Beyond chemical signatures, high-precision proper motion studies, primarily conducted with the HST, have revealed that different populations often display distinct spatial distributions and kinematic properties \citep[e.g.,][]{2009A&A...507.1393B,2023ApJ...944...58L,2025ApJ...986...80G}. 
The MPs problem involves various aspects, ranging from star formation and stellar structure and evolution of the first generation of stars in the early Universe. We will not elaborate on these details here as \rev{they are} beyond the scope of this work. We refer interested readers to \cite{universe8070359}.

The MPs phenomenon is a classic case where ``observations precede theory''. Although it has been nearly half a century since the discovery of MPs, no model currently exists that can comprehensively explain their origin and detailed observational manifestations. For the development of theories regarding MPs, we recommend the work of \cite{2022MNRAS.513.2111R}, as well as the critical assessment of these models provided by \cite{2018ARA&A..56...83B}. A significant fraction of the challenges confronting these models stems from the incomplete coverage of the relevant observational parameter space. For instance, the Milky Way contains virtually no young clusters with densities and masses comparable to those of ancient GCs. Conversely, in extragalactic starburst environments (e.g., the Magellanic Clouds), disentangling the MS members of young clusters from the surrounding field population remains technically challenging. Furthermore, scenarios invoking extreme stellar physics, such as the supermassive star hypothesis \citep{2014MNRAS.437L..21D, 2018MNRAS.478.2461G}, rely on objects that have thus far eluded direct detection. Consequently, current stellar evolution models struggle to robustly constrain the structure and evolution of these theoretical extreme stars. Although the HST possesses unrivalled capabilities in resolving MPs within crowded environments \citep{2015AJ....149...91P}, its restricted field of view (merely $162'' \times 162''$) imposes significant constraints on the accessible sample size. Consequently, the full census of the over 160 known Galactic GCs remains incompletely surveyed by HST \citep[see][]{1996AJ....112.1487H}, let alone the thousands of Galactic open clusters and extragalactic young massive clusters.

Given that the HST has been operational for over three decades, the field of MPs research currently faces the risk of diminishing telescope accessibility. Consequently, the deployment of a wide-field space telescope in the coming decade--featuring spectral coverage, filter definitions, and astrometric and photometric precision comparable to those of the HST, is critical for ensuring the continuity of this research. Such a facility would not only guarantee the acquisition of a significantly larger cluster sample with data quality commensurate with that of the HST, but also facilitate the extension of temporal baselines by leveraging historical HST observations. This capability is particularly crucial for disentangling member stars in extragalactic clusters via high-precision proper motions \citep[e.g.,][]{2023A&A...672A.161M}. The \textit{Chinese Space Station Survey Telescope} (CSST) represents precisely such a facility \citep{2025arXiv250704618C}.

The CSST is an independent optical space observatory designed to co-orbit with the China Space Station (CSS). When necessary, it can dock with the station for maintenance. The CSST employs a single primary optical system, including a $2$-metre aperture, $f/14$ off-axis three-mirror anastigmat of Cooke configuration, to collect light from observed targets. Through a folding mirror near the exit pupil, the system can switch between different focal instruments. The first-generation focal plane instruments include:

\begin{itemize}
\item Multi-band Imaging and Slit-less Spectroscopy Survey Camera (also known as the Survey Camera, SCam)
\item Multi-Channel Imager (MCI)
\item Integral Field Spectrograph (IFS)
\item Cool Planet Imaging Coronagraph (CPI-C)
\item THz Spectrometer (TS)
\end{itemize}

Previous work has demonstrated that the MCI onboard the CSST possesses capabilities for characterizing MPs comparable to those of the UVIS/WFC3 on the HST \citep{2022RAA....22i5004L}. However, the SCam is designated to receive the majority of the observing time allocation under the current CSST survey strategy. Furthermore, the conclusions drawn by \citet{2022RAA....22i5004L} were predicated primarily on theoretical isochrones derived from stellar evolution models, without accounting for realistic observational uncertainties and instrumental signatures through image-level simulations. Consequently, prior to the launch and subsequent data release of the CSST, it is imperative to rigorously evaluate the specific capability of SCam to detect and characterise MPs using high-fidelity simulations that closely mimic actual observing conditions. Adopting the image-level simulation workflow recently introduced by \cite{2026ApJ..1000..122L} for stellar streams, the primary objective of this paper is to conduct a comprehensive analysis focused on globular clusters, thereby establishing a robust observational framework and providing strategic guidance for future MP research in the CSST era.

The article is organised as follows: in \autoref{sect:bol}, we introduce the details of our method. In \autoref{sect:result} we show our main results . In \autoref{sect:conclusion} we summarise our conclusions.

\section{Bolometric Corrections for Multiple Stellar
Populations}
\label{sect:bol}
The SCam comprises $2.6 \times 10^{9}$ pixels, covering a $1.1~\mathrm{deg}^{2}$ field of view, with an image quality of $\mathrm{EE80} \leq 0.15^{\prime\prime}$ (radius containing 80\% of the PSF energy, \cite{2026RAA....26b4002B}). The focal plane is divided into seven imaging bands ($NUV$, $u$, $g$, $r$, $i$, $z$, $y$) and three slitless spectroscopy bands (GU, GV, GI), with an average spectral resolution $R \geq 200$ and wavelength coverage from $255$--$1000$~ nm. The passband response curves are shown in Figure~\ref{fig1}, and the main focal plane layout of the SCam can be obtained through \cite{2026RAA....26b4001W}, their figure 3.

This study primarily focuses on the multi-band imaging component of the survey. Table~\ref{table:1} summarises the core parameters of each filter:
(1) Central wavelength ($\lambda_c$);
(2) Full width at half maximum (FWHM);
(3) the corresponding wavelengths of 50\% of the maximum transmission curve (left $\lambda_{L50}$ and right $\lambda_{R50}$);
(4) Average transmission efficiency within the FWHM range ($T_{50}$).

To evaluate the impact of MPs on the CSST/SCam photometric system, we need to select a specific chemical mode to simulate its influence on photometric measurement. We constructed a synthetic MP model with chemical abundances mimicking those of NGC 2808. This specific target was selected because NGC 2808 serves as an archetypal Galactic GC known to host MPs characterised by substantial abundance variations, spanning from light elements (C, N, O) to significant Helium enrichment \citep{2007ApJ...661L..53P, 2015ApJ...808...51M}.  We utilised the Dartmouth Stellar Evolution Database (Dartmouth model) to generate isochrones with different He abundances \citep{2008ApJS..178...89D}.
Based on the research of \cite{2007ApJ...661L..53P} we adopt the age and metallicity of our two populations, both $t = 12~\text{Gyr}$ and $[\text{Fe}/\text{H}] = -1.0~\text{dex}$. They are different in He abundance, with $Y = 0.25$ for one population and $Y = 0.40$ for the other (hereafter 1P and 2P). We set the $[\alpha/\text{Fe}] = 0.0~\text{dex}$ as we \rev{confirmed} that it has a negligible effect on the derived isochrones. In the Dartmouth simulations, we obtained relevant stellar data, including stellar mass, $\log T_{\text{eff}}$, $\log g$, and $\log L/L_\odot$. These data, along with the specified $[\text{Fe}/\text{H}]$, were used to compute bolometric corrections through the PARSEC database, yielding absolute magnitudes in multi-band imaging for the CSST/SCam. 

We employed \textsc{SPECTRUM} (version~2.77, \cite{1994AJ....107..742G}) to compute synthetic spectra for stars along the isochrones with specific elemental abundances, thereby determining the loci for stellar populations. Given a stellar atmosphere model and specific input parameters, \textsc{SPECTRUM} calculates synthetic stellar spectra. The software reliably generates synthetic spectra for B-type to mid-M-type stars. For the opacity calculations, we utilised the master line list \textsc{lukeall2} provided with the package. This list is a compilation of atomic and molecular data derived primarily from the NIST database and Kurucz line lists, empirically adjusted to match the solar spectrum. including more than $1.6 \times 10^8$ absorption lines that cover a wavelength range from $900$~\AA \;to $40000$~\AA. The atmospheric models used as input were computed with the \textsc{ATLAS9} model atmosphere program developed by \cite{1993sssp.book.....K}, with parameters ($[\mathrm{Fe}/\mathrm{H}]$, $\log g$, $T_{\mathrm{eff}}$) derived from the base isochrone from Dartmouth.

We generated stars with different elemental abundances. With $[\mathrm{Fe}/\mathrm{H}] = -1.0$ and constant total C+N+O abundance for all stars, we set $\delta[\mathrm{N}/\mathrm{Fe}] = +1.0~\mathrm{dex}$ and $\delta[\mathrm{O}/\mathrm{Fe}] = \delta[\mathrm{C}/\mathrm{Fe}] = -0.5~\mathrm{dex}$ for all 2P stars. According to \cite{2022RAA....22i5004L}, elements such as Na and Mg have a negligible effect on wide-band filter magnitudes and are therefore not considered in this study. The synthetic stellar spectra provide the absolute flux at each wavelength.

The absolute magnitude of a star belonging to the chemically enriched population (that is, 2P) in the i-th filter is calculated using Equation \ref{eq1}:

   \begin{equation}
    \label{eq1}
      M_i = -2.5\log[\frac{{\int_{\lambda_1}^{\lambda_2} \lambda S_{\lambda,i} f_\lambda d\lambda}} {\int_{\lambda_1}^{\lambda_2} \lambda S_{\lambda,i} f_\lambda^0 d\lambda} ]+M_{i,0}\,,
   \end{equation}

where $f_\lambda$ is the flux at wavelength $\lambda$ for a chemically enriched star at a distance of 10 pc, $f^0_\lambda$ denotes the flux of a reference star, which serves as a proxy for the stellar population synthesised using the default stellar chemical abundance table, $S_{\lambda,i}$ is the transmission of the CSST filter at wavelength $\lambda$, and $M_i$ and $M_{i,0}$ are the absolute magnitudes of the chemically enriched and normal stars, respectively. Equation \ref{eq1} is valid for the present-day photon-counting devices (CCDs or IR arrays).

Although the spectral synthesis code \textsc{SPECTRUM 2.77} enables the generation of synthetic spectra with varying C, N, and O abundances, the standard Dartmouth stellar evolution models are limited to discrete helium mass fractions ($Y$ = 0.25, 0.33 and 0.40) \citep{2008ApJS..178...89D}. Consequently, to more rigorously constrain the capability of the CSST to resolve MPs, we employed the \texttt{PGPUC} models \citep{2013A&A...553A..62V} to construct a denser grid, thereby allowing for the exploration of a much finer parameter space in chemical abundances. We generated \texttt{PGPUC} isochrones for seven stellar populations with an evolutionary age of $12~\mathrm{Gyr}$ and helium abundance $Y$ ranging from $0.25$ to $0.37$ in steps of $0.02$. For each population, the relationship between the variations in C, N and O and $Y$ is shown in Table~\ref{tab:abundance_change}, matching the pattern of variation of Dartmouth models at $Y = 0.40$. We fix $[\mathrm{Fe}/\mathrm{H}] = -1.0$ and calculate the corresponding metallicity $Z$ for each $Y$. To cover more GCs, we extended the distance of the clusters generated with \texttt{PGPUC} to 20 kpc. Similarly, we used \texttt{SPECTRUM} to synthesise stellar spectra, which were then convolved with the CSST/SCam filter transmission curves to obtain multi-band magnitudes.

\begin{table}[h!]
\caption{Information of filters designed for the CSST/SCam}  
\label{table:1}    
\centering  
\begin{tabular}{c c c c c c} 
\hline\hline               
Filter & $\lambda_c$ & FWHM & $\lambda_{L50}$ & $\lambda_{R50}$ & $T_{50}$ \\ 
 & (nm) & (nm) & (nm) & (nm) & \\ 
\hline                      
    NUV & 287.1 & 67.9 & 253.1 & 321.1 & 0.196 \\    
    u   & 360.2 & 83.0 & 318.7 & 401.7 & 0.292 \\    
    g   & 476.8 & 151.0 & 401.3 & 552.3 & 0.579 \\    
    r   & 619.1 & 144.6 & 546.8 & 691.4 & 0.642 \\    
    i   & 764.3 & 154.2 & 687.2 & 841.4 & 0.639 \\    
    z   & 890.0 & 107.5 & 836.3 & 943.8 & 0.465 \\    
    y   & 954.0 & 57.8 & 925.1 & 982.9 & 0.242 \\    
\hline                                  
\end{tabular}
\vspace{0.2cm}
\footnotesize
\begin{minipage}{\linewidth}
\textbf{Notes:} $\lambda_c$ : Central wavelength. FWHM : Full width at half maximum. $\lambda_{L50}$ and $\lambda_{R50}$ : The wavelengths at which the transmission drops to 50\% on the left and right sides respectively. $T_{50}$ : Average transmission efficiency within the FWHM range.
\end{minipage}
\end{table}

\begin{table}[ht]
\centering
\caption{Adopted parameters in PGPUC model}
\label{tab:abundance_change}
\begin{tabular}{cccc}
\hline
Y & $\delta \mathrm{C}$ & $\delta \mathrm{N}$ & $\delta \mathrm{O}$ \\
& [dex] & [dex] & [dex] \\
\hline
0.27 & $-0.04$ & $+0.32$ & $-0.04$ \\
0.29 & $-0.09$ & $+0.51$ & $-0.09$ \\
0.31 & $-0.14$ & $+0.64$ & $-0.14$ \\
0.33 & $-0.21$ & $+0.73$ & $-0.21$ \\
0.35 & $-0.28$ & $+0.81$ & $-0.28$ \\
0.37 & $-0.36$ & $+0.88$ & $-0.36$ \\
\hline
\end{tabular}
\vspace{0.2cm}
\footnotesize
\begin{minipage}{\linewidth}
\textbf{Notes:} relationship between Y and $\delta C,\delta N $ and $\delta O$, which matches the variation pattern of the Dartmouth model when $Y = 0.40$. 
\end{minipage}
\end{table}
   
\begin{figure}[h!]
   \centering
    \includegraphics[width=\hsize]{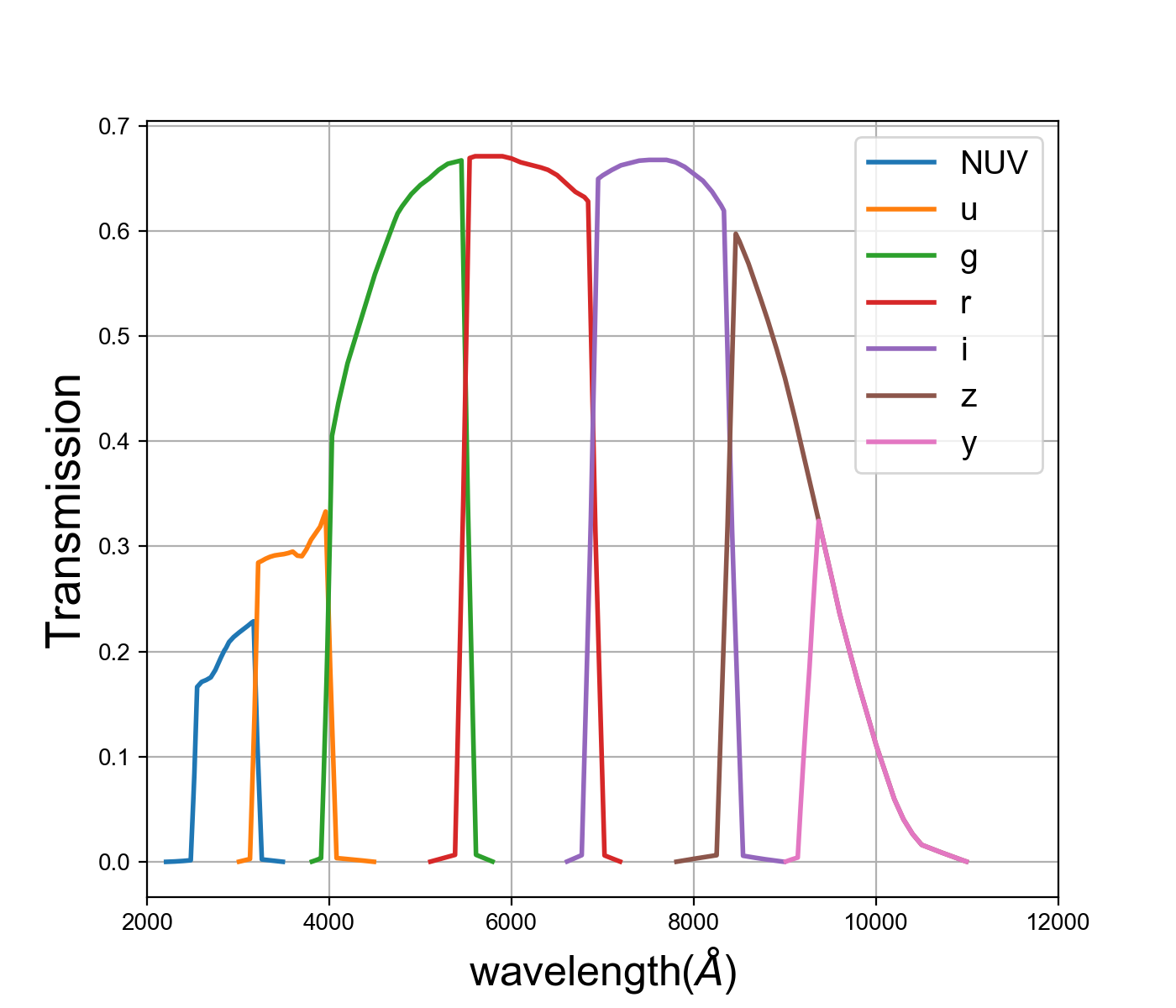}
    \caption{The total transmission curves for CSST/SCam filters, with detector quantum efficiency being considered.}
    \label{fig1}
    \end{figure}

\section{Main results}
\label{sect:result}

In this section, we first present the theoretical loci of stellar populations characterised by varying chemical abundances within the CSST/SCam CMDs, explicitly excluding photometric uncertainties and instrumental effects. Subsequently, building upon these theoretical baselines, we conduct image simulations and present the resulting photometry, which incorporates realistic simulated errors and instrumental signatures.

\subsection{Helium Variation}

In Figure~\ref{fig2} , we present the loci for 1P (normal helium abundance, $Y = 0.25$, red) and 2P ($Y = 0.40$, black) stars (left panel). Here 1P and 2P stars are only different in helium abundance. The upper right panel shows the spectra of two giant stars at the same evolutionary stage but with different helium abundances, while the lower right panel displays their flux ratio in the form of $-2.5 \log(F_{normal} / F_{enrich})$, where $F_{normal}$ and $F_{enrich}$ represent the fluxes of the 1P and 2P stars, respectively. This ratio is equivalent to a magnitude difference spectrum. As mentioned earlier, helium-rich stars appear bluer and brighter than stars with normal chemical abundances.

Our results demonstrate that the MS and RGB sequences composed of Helium-enriched stars are systematically bluer than those of populations with standard Helium abundances. This is physically expected because at constant metallicity, helium enrichment leads to an increase in the mean molecular weight $\mu$ and a decrease in the opacity $\kappa$. According to the stellar structure equations, an increase in $\mu$ results in a higher core temperature and accelerated hydrogen burning, while a decrease in $\kappa$ leads to a reduced temperature gradient and an increase in both the effective temperature and luminosity $L$. However, the sequences intersect near the MS Turn-off (MSTO) and SGB, where the Helium-enriched population appears marginally fainter than the standard population. This behaviour is attributed to the enhanced nuclear burning efficiency induced by Helium enrichment, which accelerates stellar evolution. Consequently, at a fixed isochronal age, Helium-enriched stars situated at the MSTO and SGB possess systematically lower masses than their standard counterparts. Since stellar mass is the primary determinant of luminosity during these phases, the Helium-enriched population manifests as less luminous. As the stars evolve onto the RGB, however, the intrinsic influence of Helium abundance on stellar structure reasserts its dominance.

   \begin{figure*}[h!]
        \centering
        \begin{minipage}{0.47\hsize}
            \centering
             \includegraphics[width=\linewidth,height=0.5\linewidth]{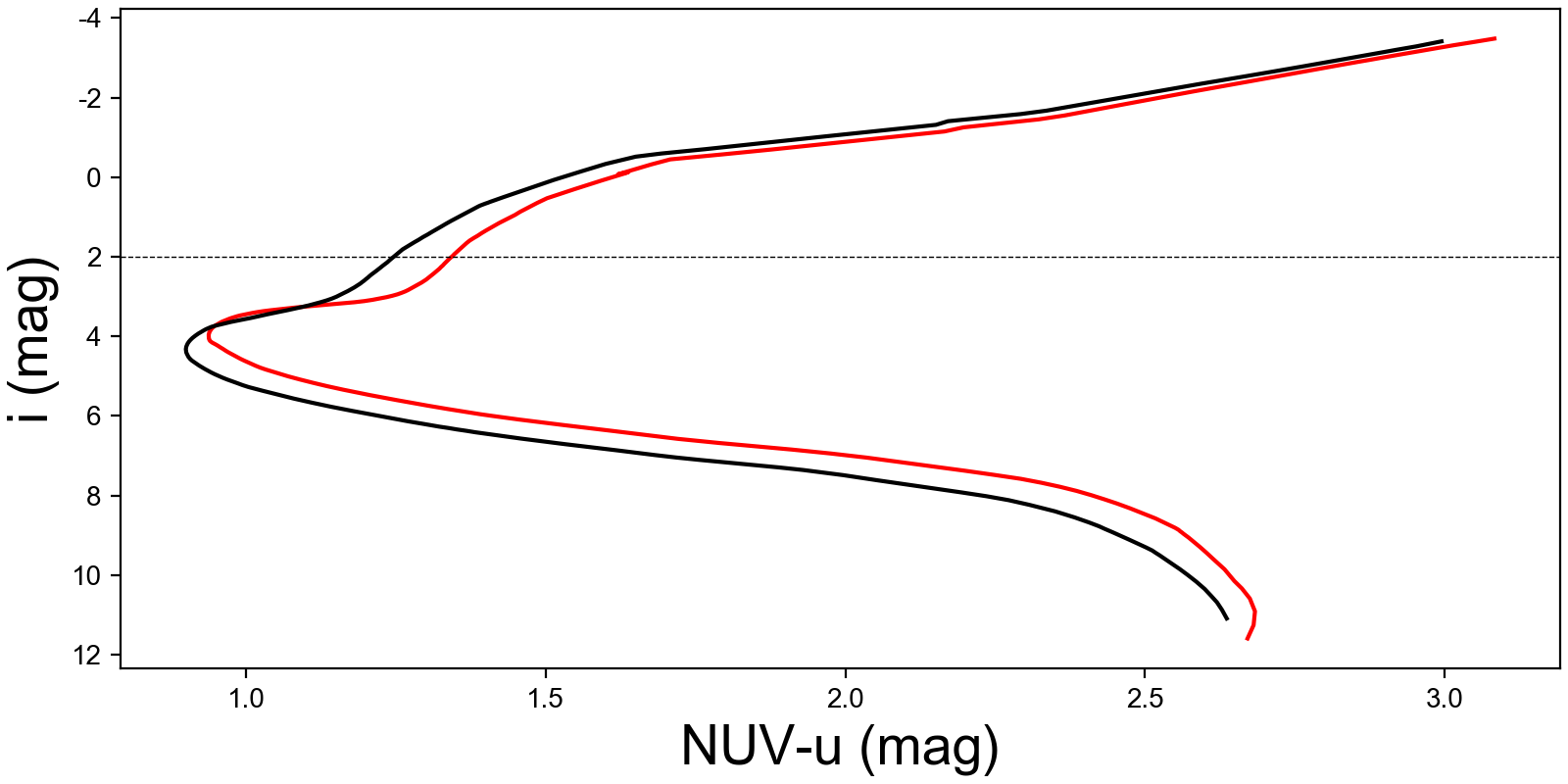}
             \vspace{0.5\baselineskip}
             \includegraphics[width=\linewidth,height=0.5\linewidth]{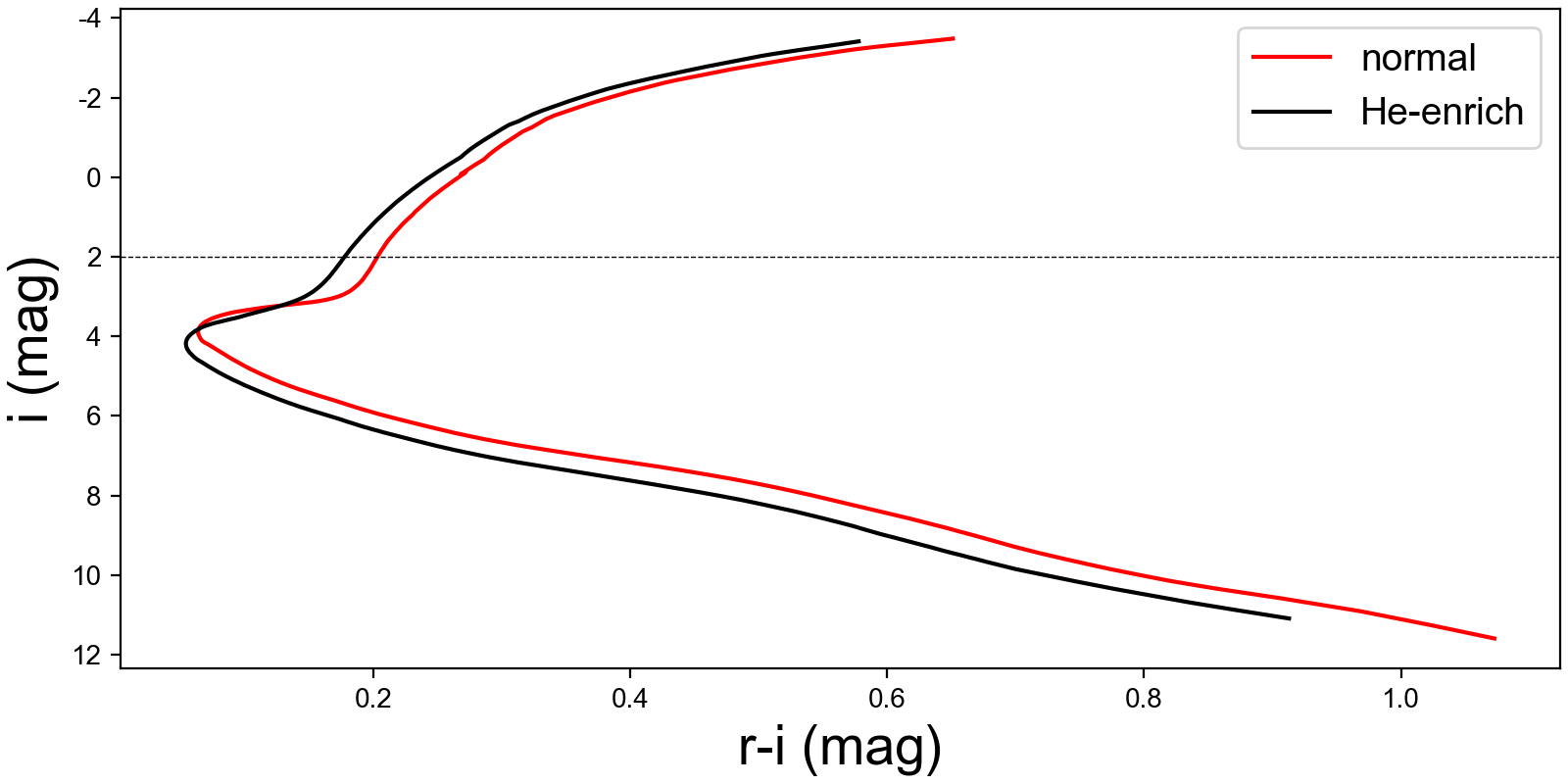}
        \end{minipage}
        \begin{minipage}{0.47\hsize}
            \centering
             \includegraphics[width=\linewidth,height=0.5\linewidth]{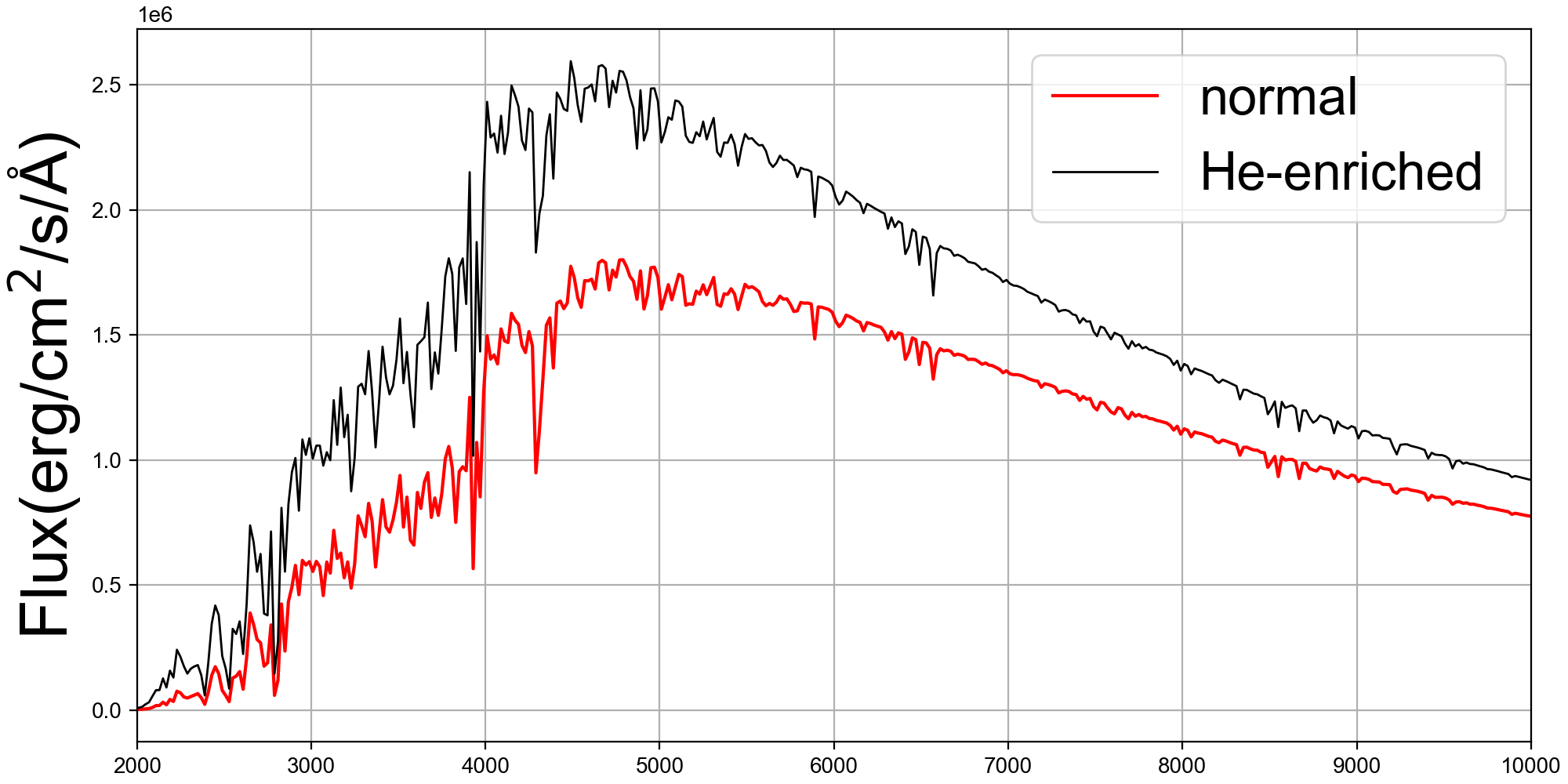}
             \vspace{0.5\baselineskip}
             \includegraphics[width=\linewidth,height=0.5\linewidth]{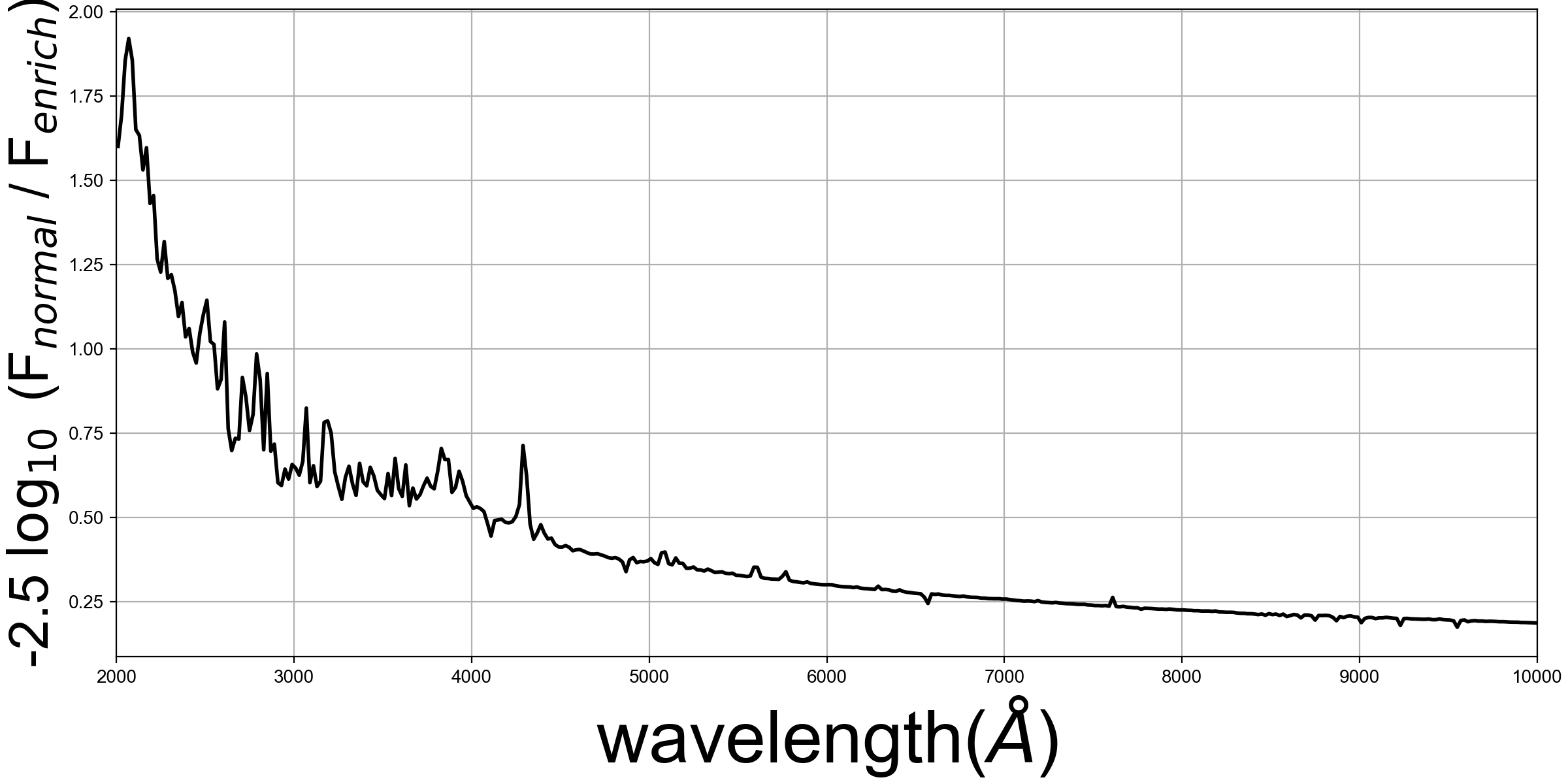}
        \end{minipage}
        \caption{Upper left panel: Loci of stellar populations with $Y=0.25$ (red) and $Y=0.40$ (black) in the $NUV - u$ vs. $i$ CMD. Lower left panel: Same loci as the upper left panel, but plotted in the $r - i$ vs. $i$ CMD. Upper right panel: Spectra of two MS stars (red: $Y=0.25$; black: $Y=0.40$) that share the same $i$-band magnitude, as marked by the dotted line in the left panel. Lower right panel: Magnitude difference spectra between the two stars. As the abundance of helium increases, stars become hotter. At the same i-band magnitude, helium-rich stars have slightly smaller masses and higher spectral flux per unit area.}
        \ContinuedFloat 
        \label{fig2}
    \end{figure*}

\subsection{Carbon, Nitrogen and Oxygen Variations}
 Figure~\ref{fig3} displays the loci of two stellar populations in the CMDs, differing solely in their CNO abundances. Specifically, the 2P population is enriched in N by 1.0 dex relative to the 1P population, whereas C and O are both depleted by 0.5 dex. The top-left panel presents the $(NUV - u)$ versus $i$ CMD, while the bottom-left panel shows the optical $(r - i)$ versus $i$ CMD. We observe that the 2P population is systematically bluer than the 1P population in the $(NUV - u)$ colour index. In contrast, the loci of the 1P and 2P populations nearly coincide in the $(r - i)$ versus $i$ CMD. This discrepancy can be understood from the right panels of Figure~\ref{fig3}. Analogously to Figure~\ref{fig2}, the top-right panel displays the spectra of two red giants from the 1P and 2P populations at the same evolutionary stage. The bottom-right panel presents their `magnitude difference' spectrum, defined as $-2.5 \log(F_{normal} / F_{enrich})$ as a function of wavelength. In the wavelength range of $<3000$\,\AA, the continuum of red giants is extremely weak, and the stellar flux is predominantly governed by OH molecular absorption bands. Due to the O-depletion in 2P stars relative to the 1P population, the flux of 2P stars is enhanced in the $NUV$ band, leading to a decrease in the $NUV$ magnitude. Conversely, across a broad region centred at $\sim3370$\,\AA, the stars are subject to deep NH molecular absorption bands. As a result of the N-enrichment in 2P stars, this absorption is intensified, suppressing the flux in the $u$ band and causing the $u$-band magnitude to increase relative to the 1P stars. Consequently, the $(NUV - u)$ colour index of the 2P stars becomes bluer compared to their 1P counterparts. The $r$ and $i$ bands, however, are dominated by the continuum of RGB stars, where the absorption effects induced by variations in C, N, and O abundances are negligible. Consequently, the positions of the 1P and 2P stars are virtually identical in the $(r - i)$ versus $i$ CMD. \footnote{In fact, a distinct absorption feature dominated by the CH molecule exists at $\sim 4300$\,\AA, which falls within the coverage of the $g$-band filter designed for the CSST/SCam. However, the stellar continuum in this wavelength range remains sufficiently intense to overwhelm the minor variations induced by the CH absorption. Consequently, the $(g - i)$ colour index is also insensitive to MPs.
}

   \begin{figure*}[h!]
        \centering
        \begin{minipage}{0.47\hsize}
            \centering
             \includegraphics[width=\linewidth,height=0.5\linewidth]{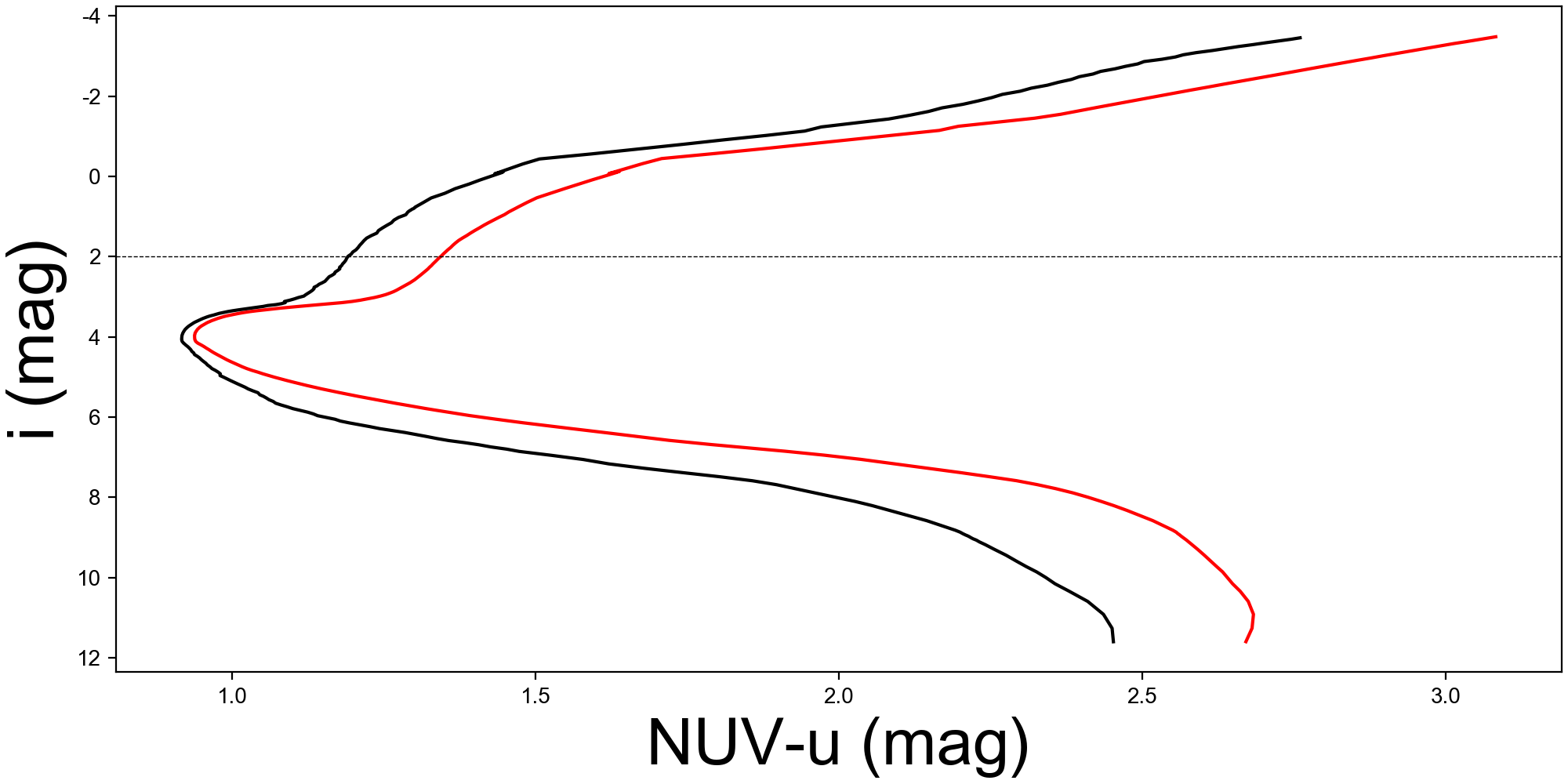}
             \vspace{0.5\baselineskip}
             \includegraphics[width=\linewidth,height=0.5\linewidth]{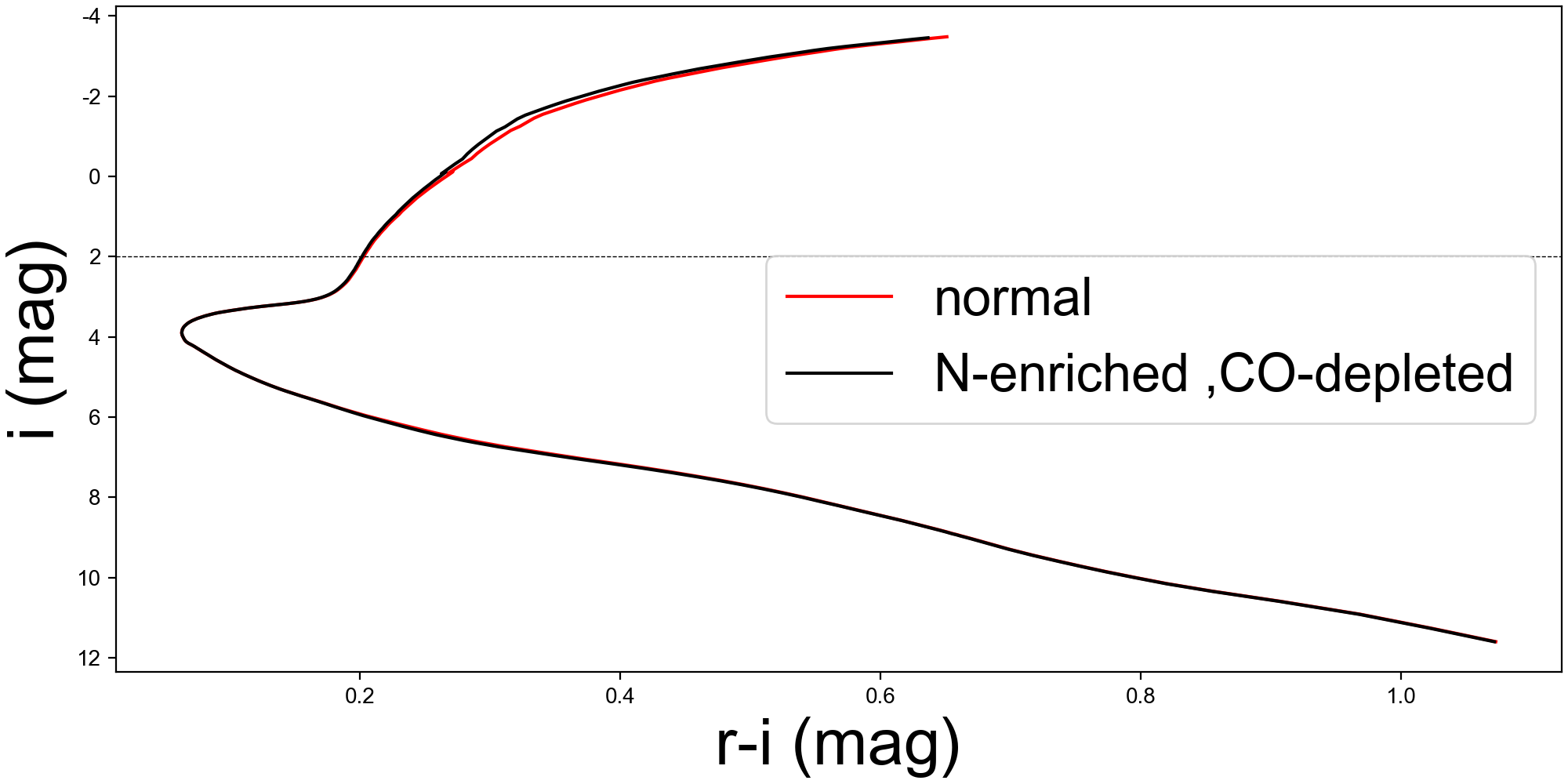}
        \end{minipage}
        \begin{minipage}{0.47\hsize}
            \centering
             \includegraphics[width=\linewidth,height=0.5\linewidth]{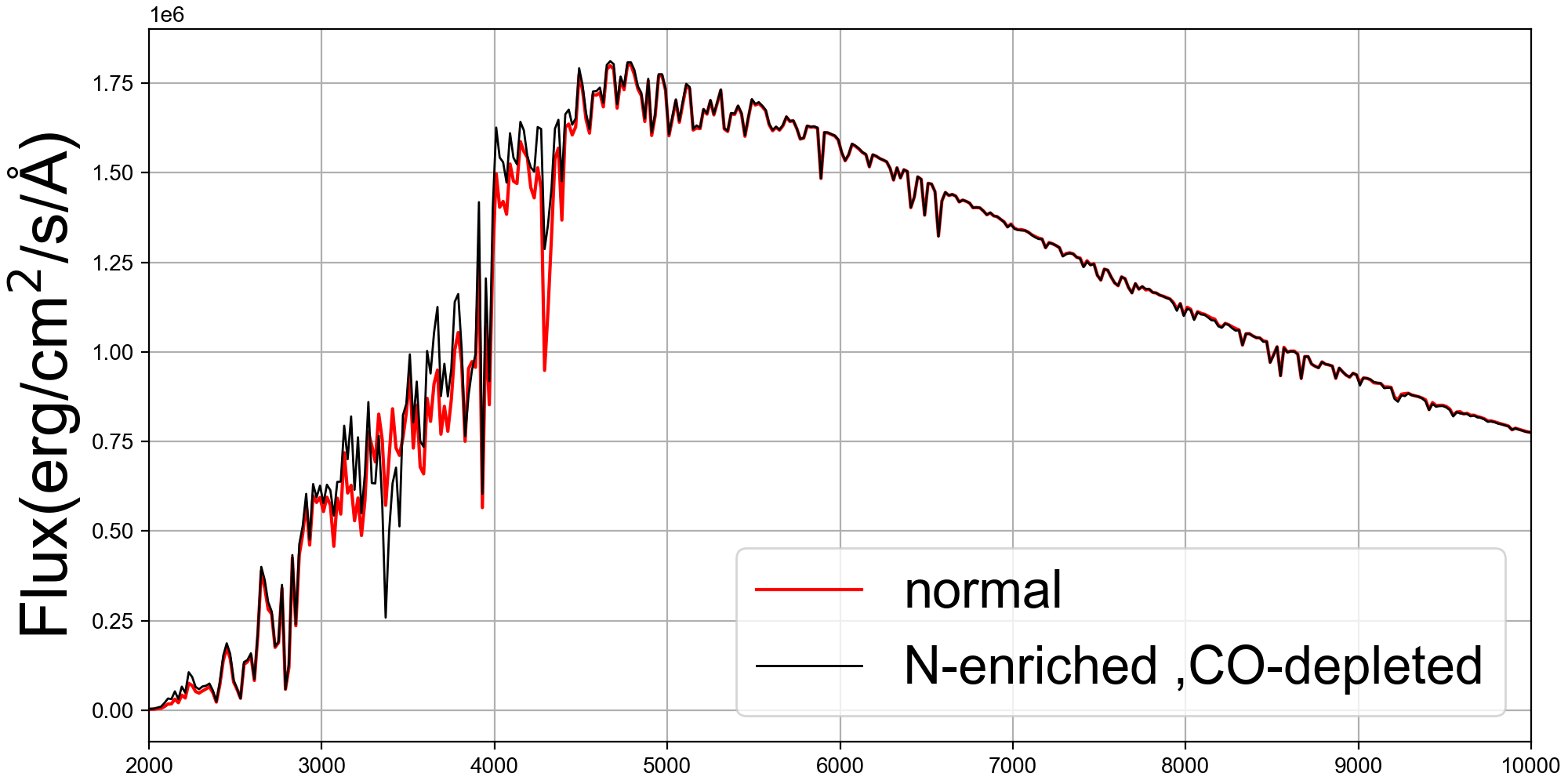}
             \vspace{0.5\baselineskip}
             \includegraphics[width=\linewidth,height=0.5\linewidth]{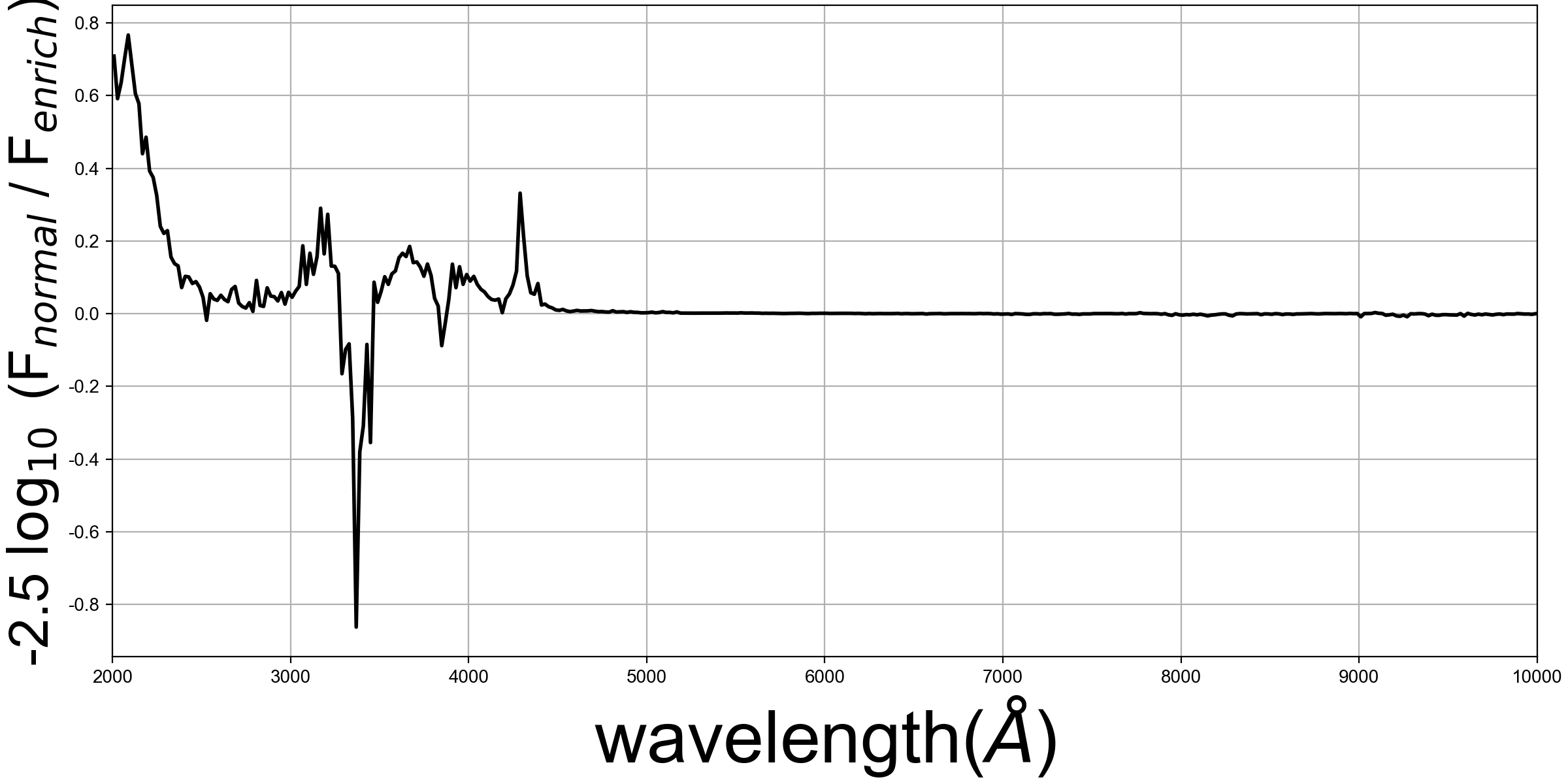}
        \end{minipage}
        \caption{Upper left panel: Loci of stellar populations with normal chemical abundance (red) and with CNO abundance variations (black) in the $NUV - u$ vs. $i$ CMD, where both populations have a helium abundance of $Y=0.25$. Lower left panel: Same loci as the upper left panel, but plotted in the $r - i$ vs. $i$ CMD. Upper right panel: Spectra of two giant stars (red: without CNO variations; black: with CNO variations) sharing the same $i$-band magnitude, as marked by the dotted line in the upper left panel. Lower right panel: Magnitude difference spectra between the two stars.}
        \ContinuedFloat 
        \label{fig3}%
    \end{figure*}

\subsection{He$+$CNO Variations}
\begin{figure}[]
\centering
\includegraphics[width=\hsize]{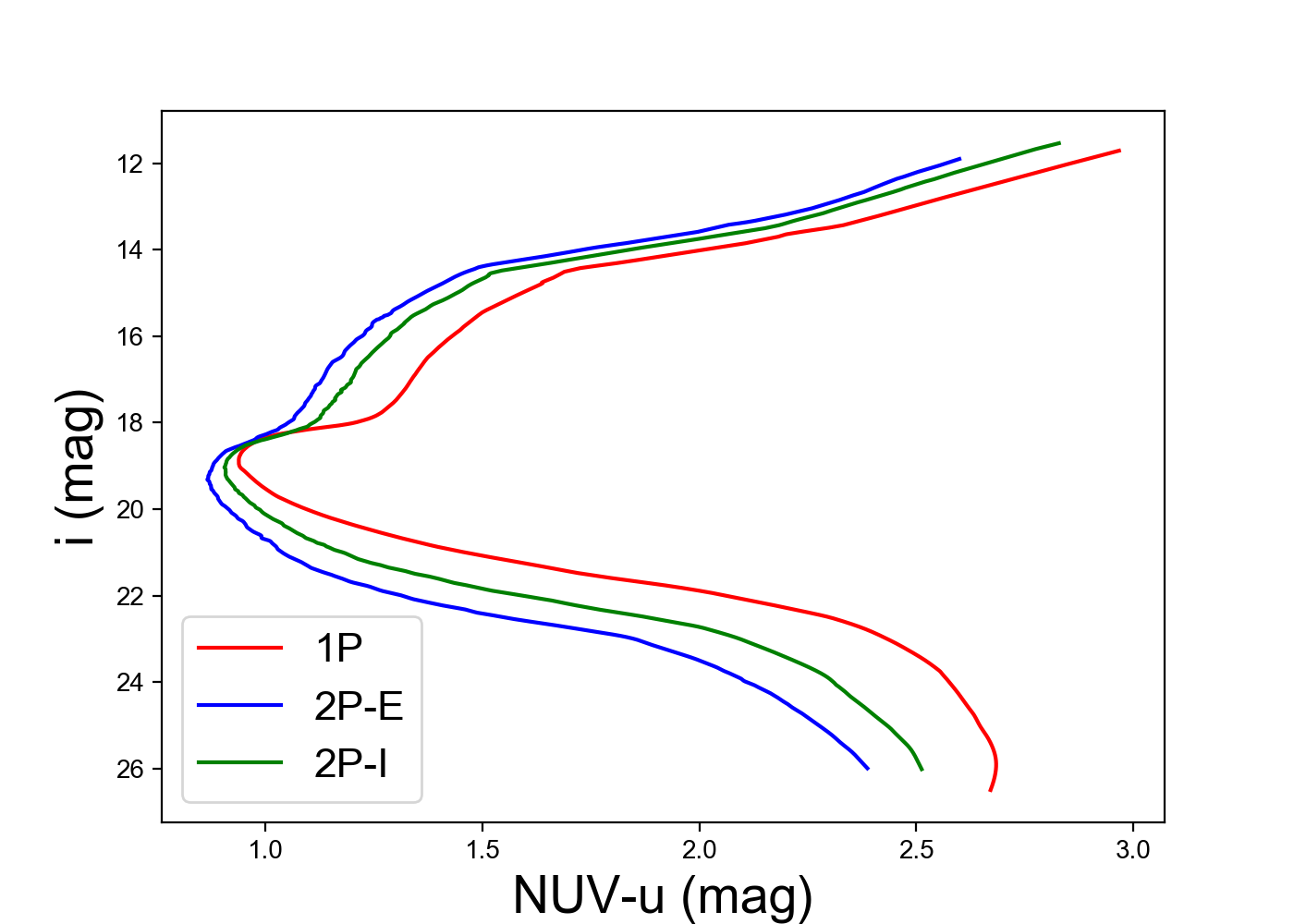}
  \caption{Similar to the left panel of Figure 2, this panel displays the distributions of stellar populations 1P, 2P-E, and 2P-I (see the main text for detailed definitions).}
     \label{fig4}
\end{figure}

We consider a 2P stellar population that /simultaneously exhibits variations in helium and CNO chemical abundances. One subset has the most extreme abundance variations (Extreme, or 2P-E, blue: $Y = 0.40$, $\delta[\mathrm{C}/\mathrm{Fe}] = -0.5,\mathrm{dex}$, $\delta[\mathrm{O}/\mathrm{Fe}] = -0.5,\mathrm{dex}$, $\delta[\mathrm{N}/\mathrm{Fe}] = +1.0,\mathrm{dex}$), while another subset has relative moderate chemical dispersion (Intermediate, or 2P-I, green: $Y = 0.33$, $\delta[\mathrm{C}/\mathrm{Fe}] = -0.3,\mathrm{dex}$, $\delta[\mathrm{O}/\mathrm{Fe}] = -0.3,\mathrm{dex}$, $\delta[\mathrm{N}/\mathrm{Fe}] = +1.0,\mathrm{dex}$), along with a chemically normal 1P (red) stellar population. Their CMD is shown in Fig~\ref{fig4}.

Tables~\ref{tab:rgb} and \ref{tab:ms} list the colour differences between the enriched stars in the most extreme scenario (population 2P-E) and their corresponding 1P counterparts. Specifically, Table~\ref{tab:rgb} represents red giants at an absolute magnitude of $i = 1.5$\,mag (G0--G5 type), while Table~\ref{tab:ms} focuses on MS dwarfs at $i = 7$\,mag (K5--K7 type). We examined various combinations of colour indices, including the $NUV$ and $ugri$ bands. We find that for stars differing solely in helium abundance, the colour difference is maximised by forming an index using filters with the largest possible separation in central wavelength (e.g., $NUV - i$). In contrast, for stars subject to CNO abundance variations (characterised by N-enhancement and C/O-depletion), the $(NUV - u)$ index proves most effective in maximising the colour difference. The underlying physical reason is evident: helium variations primarily impact the stellar continuum by modifying the effective temperature, meaning the colour difference naturally scales with the wavelength baseline of the filters. Conversely, CNO variations alter local molecular absorption features; the most pronounced absorption differences appear in the $NUV$ and $u$ bands and exhibit an anti-correlated behaviour. Consequently, $(NUV - u)$ is the optimal colour index for detecting MP phenomena driven by CNO variations.

\begin{table}[h]
\centering
\caption{colour differences for example red giants belong to 1P and 2P}
\label{tab:rgb}
\begin{tabular}{lrrr}
\hline
colour difference & He & CNO & He$+$CNO \\
 & (mag) & (mag) & (mag) \\
\hline
$\Delta({\mathrm{NUV}} - {u})$ & -0.088 & -0.157 & -0.203 \\
$\Delta({\mathrm{NUV}} - {g})$ & -0.192 & -0.057 & -0.243 \\
$\Delta({\mathrm{NUV}} - {r})$ & -0.243 & -0.064 & -0.297 \\
$\Delta({\mathrm{NUV}} - {i})$ & -0.267 & -0.065 & -0.321 \\
$\Delta({u} - {g})$ & -0.106 & 0.100 & -0.040 \\
$\Delta({u} - {r})$ & -0.155 & 0.094 & -0.093 \\
$\Delta({u} - {i})$ & -0.179 & 0.092 & -0.118 \\
$\Delta({g} - {r})$ & -0.050 & -0.007 & -0.054 \\
$\Delta({g} - {i})$ & -0.0738 & -0.0086 & -0.0782 \\
$\Delta({r} - {i})$ & -0.0237 & -0.0017 & -0.0241 \\
\hline
\end{tabular}
\vspace{0.2cm}
\footnotesize
\begin{minipage}{\linewidth}
\textbf{Notes:} This table presents the colour differences between chemically enriched (2P) and normal (1P) giants with an absolute $i$-band magnitude of $i = 1.5$\,mag. Columns 2, 3, and 4 correspond to three cases: (1) helium enrichment only ($Y=0.40$ vs.\ $Y=0.25$); (2) C, N, and O abundance variations only; and (3) combined He$+$CNO variations. The colour differences are calculated as $(A-B)_{\text{2P}} - (A-B)_{\text{1P}}$, where $A$ and $B$ denote different photometric bands.
\end{minipage}
\end{table}

\begin{table}[h]
\centering
\caption{colour differences for example dwarfs belong to 1P and 2P}
\label{tab:ms}
\begin{tabular}{lrrr}
\hline
colour difference & He & CNO & He$+$CNO \\
 & (mag) & (mag) & (mag) \\
\hline
$\Delta({\mathrm{NUV}} - {u})$ & -0.317 & -0.456 & -0.734 \\
$\Delta({\mathrm{NUV}} - {g})$ & -0.630 & -0.305 & -0.877 \\
$\Delta({\mathrm{NUV}} - {r})$ & -0.788 & -0.308 & -1.046 \\
$\Delta({\mathrm{NUV}} - {i})$ & -0.874 & -0.310 & -1.132 \\
$\Delta({u} - {g})$ & -0.314 & 0.148 & -0.142 \\
$\Delta({u} - {r})$ & -0.477 & 0.147 & -0.315 \\
$\Delta({u} - {i})$ & -0.557 & 0.144 & -0.399 \\
$\Delta({g} - {r})$ & -0.1639 & -0.0015 & -0.1720 \\
$\Delta({g} - {i})$ & -0.2436 & -0.0059 & -0.2570 \\
$\Delta({r} - {i})$ & -0.0796 & -0.0044 & -0.0850 \\
\hline
\end{tabular}
\vspace{0.2cm}
\footnotesize
\begin{minipage}{\linewidth}
\textbf{Notes:} Same as Table 2, but for dwarf stars with absolute $i$-band magnitude $i = 7$\,mag.
\end{minipage}
\end{table}

\subsection{Mock Observations}

\begin{figure*}[h!]
    \centering
    \includegraphics[width=\hsize]{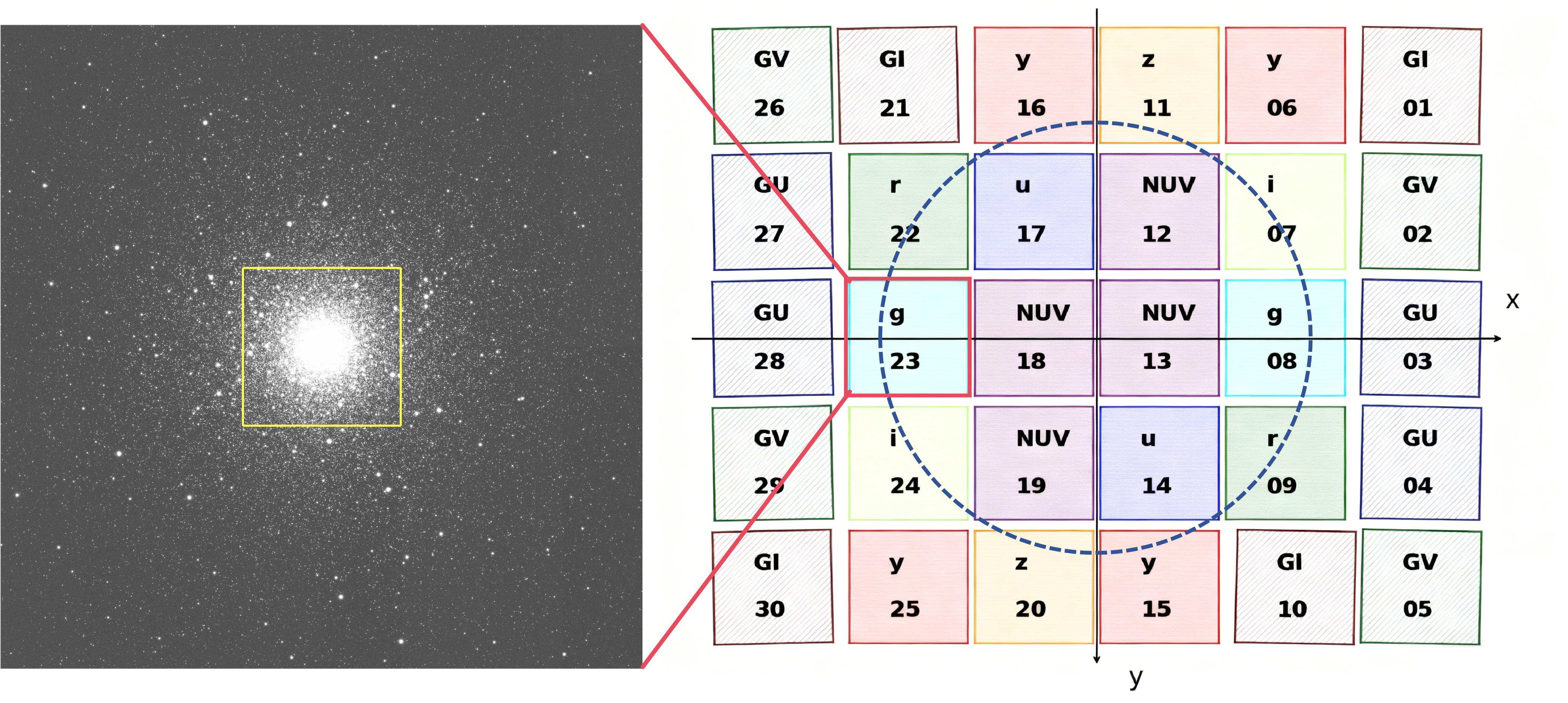}
      \caption{Left panel: Simulated cluster image generated via \texttt{CSST-msc-sim}; the FoV of this panel is $11' \times 11'$, matching the FoV of a single CCD of the CSST/SCam (shown in the right panel). The yellow square denotes the FoV of the HST UVIS/WFC3 instrument ($162'' \times 162''$). Right panel: This panel shows the segmented focal plane of the CSST/SCam (individual detector regions are labeled with passbands and identifiers). The blue dashed circle indicates the tidal radius of our simulated cluster, which corresponds to that of the GC NGC 2808.}
         \label{fig5}
    \end{figure*}

The 1P and 2P loci described above are derived solely from theoretical calculations based on isochrones and model spectra, and do not account for realistic observational effects. To address this, we utilise the virtual simulation pipeline developed internally by the CSST working group (\texttt{CSST-msc-sim}) to convert these theoretical data into mock observations. This enables us to assess the expected appearance of MPs in the CMDs under actual CSST observing conditions. Based on the Kroupa initial mass function (IMF; \cite{2001MNRAS.322..231K}), we generated 400,000 stars, equally split between 1P and 2P populations, adopting this simplified ratio for the purposes of the simulations. The colours and magnitudes of the stars were interpolated from isochrones with the appropriate chemical compositions. Using this new isochrone and the synthetic spectra, we generated two synthetic populations (1P and 2P). The positions of these synthetic stars are generated using a King model, configured with cluster parameters similar to NGC 2808: a core radius of 0.004 degrees and a tidal radius of 0.253 degrees\citep{2010arXiv1012.3224H}. In Figure~\ref{fig5}, we present the simulated cluster images generated based on \texttt{CSST-msc-sim}. We also label the tidal radius of the cluster, as well as the fields of view HST UVIS/WFC3. The field of view (11$' \times 11'$) shown in Figure~\ref{fig5} corresponds to that of a single channel of the CSST/SCam. The tidal radius (0.253$^\circ$) of the simulated GC (corresponding to NGC 2808) exceeds the coverage of a single channel, yet it is clearly smaller than the field of view (FoV) of the entire focal plane. Including \rev{NGC 2808}, over $93\%$ (according to \cite{2010arXiv1012.3224H}) of the currently known globular clusters have tidal radius angular diameters smaller than the FoV of a single pointing of CSST. The apparent magnitudes in each band were derived by adding a distance modulus of 14.91 mag ($\sim$9.6 kpc, which is the typical distance for the GC NGC 2808\citep{2010arXiv1012.3224H}) to the absolute magnitudes. This resulted in an artificial stellar catalogue containing celestial coordinates, apparent magnitudes in various bands, and synthetic spectra for each star.

    \begin{figure}[h!]
    \centering
    \includegraphics[width=\hsize]{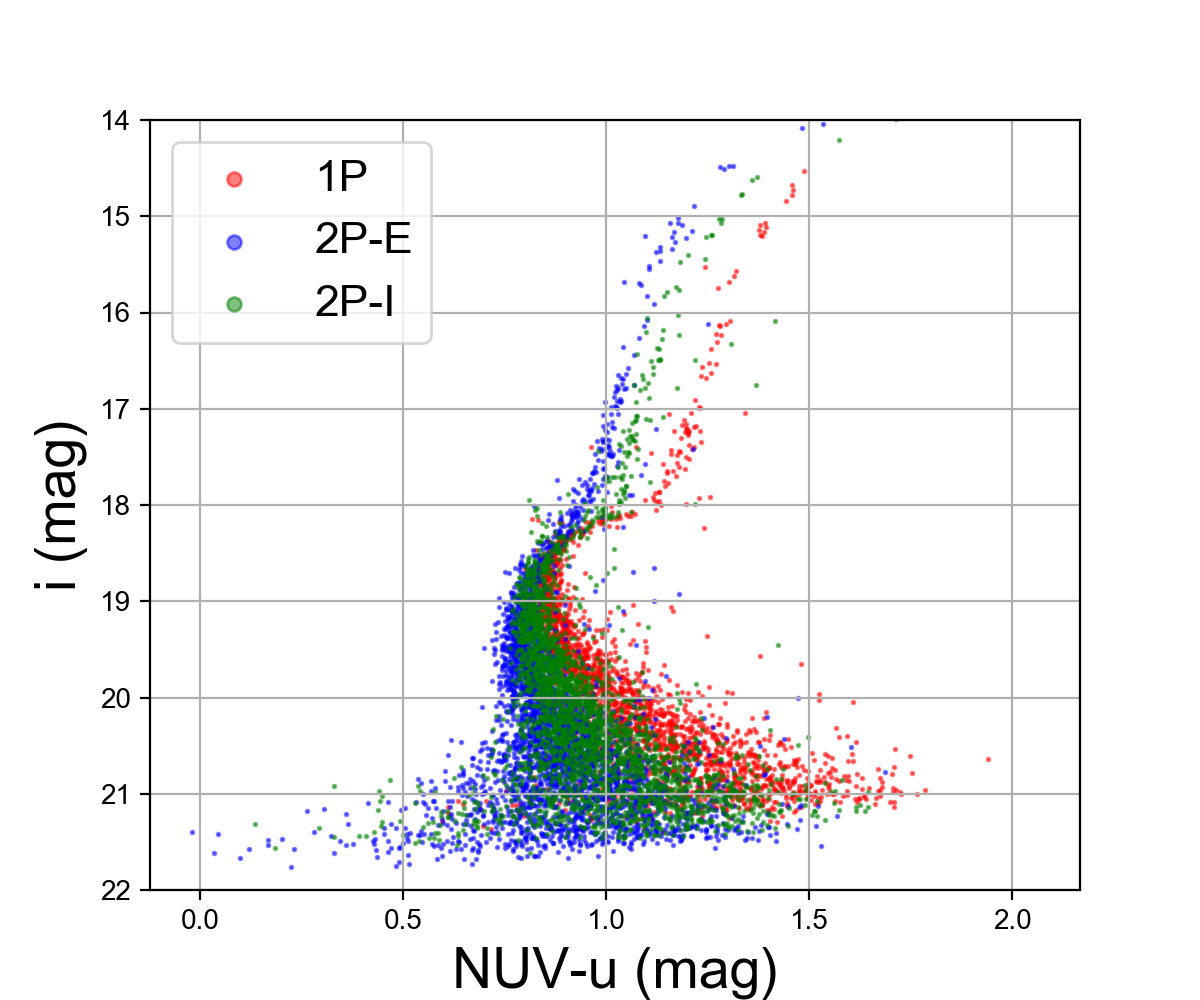}
      \caption{Figure~\ref{fig4} presents the distributions of three stellar populations (with different chemical abundances) in the CMD, as derived from mock observations.}
         \label{fig6}
    \end{figure}

    \begin{figure}[h!]
    \centering
    \includegraphics[width=\hsize]{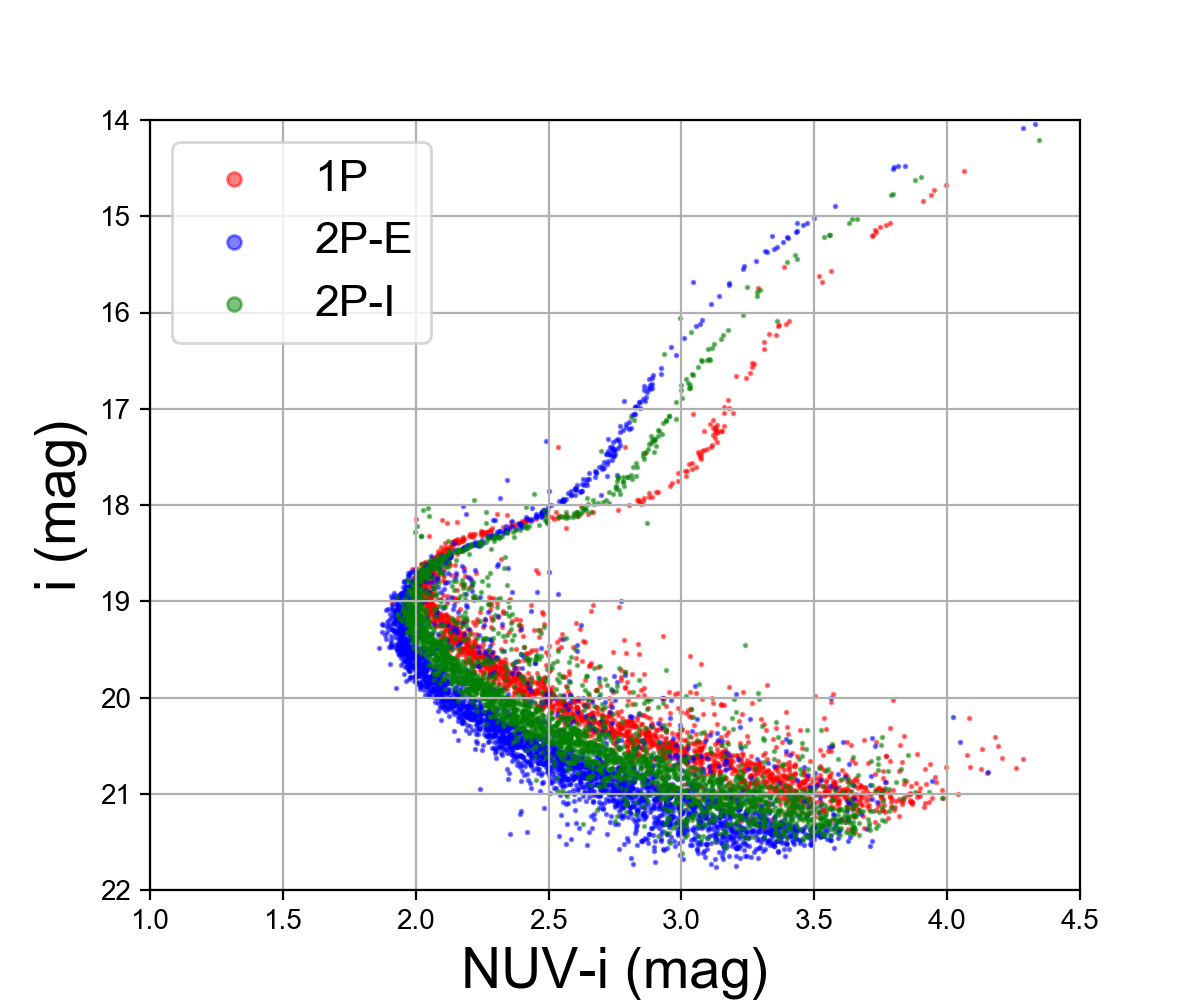}
      \caption{Same as fig~\ref{fig6}, but use $NUV-i$ as the horizontal axis.}
         \label{fig_cmdd_n-i}
    \end{figure}

    \begin{figure}[h!]
    \centering
    \includegraphics[width=\hsize]{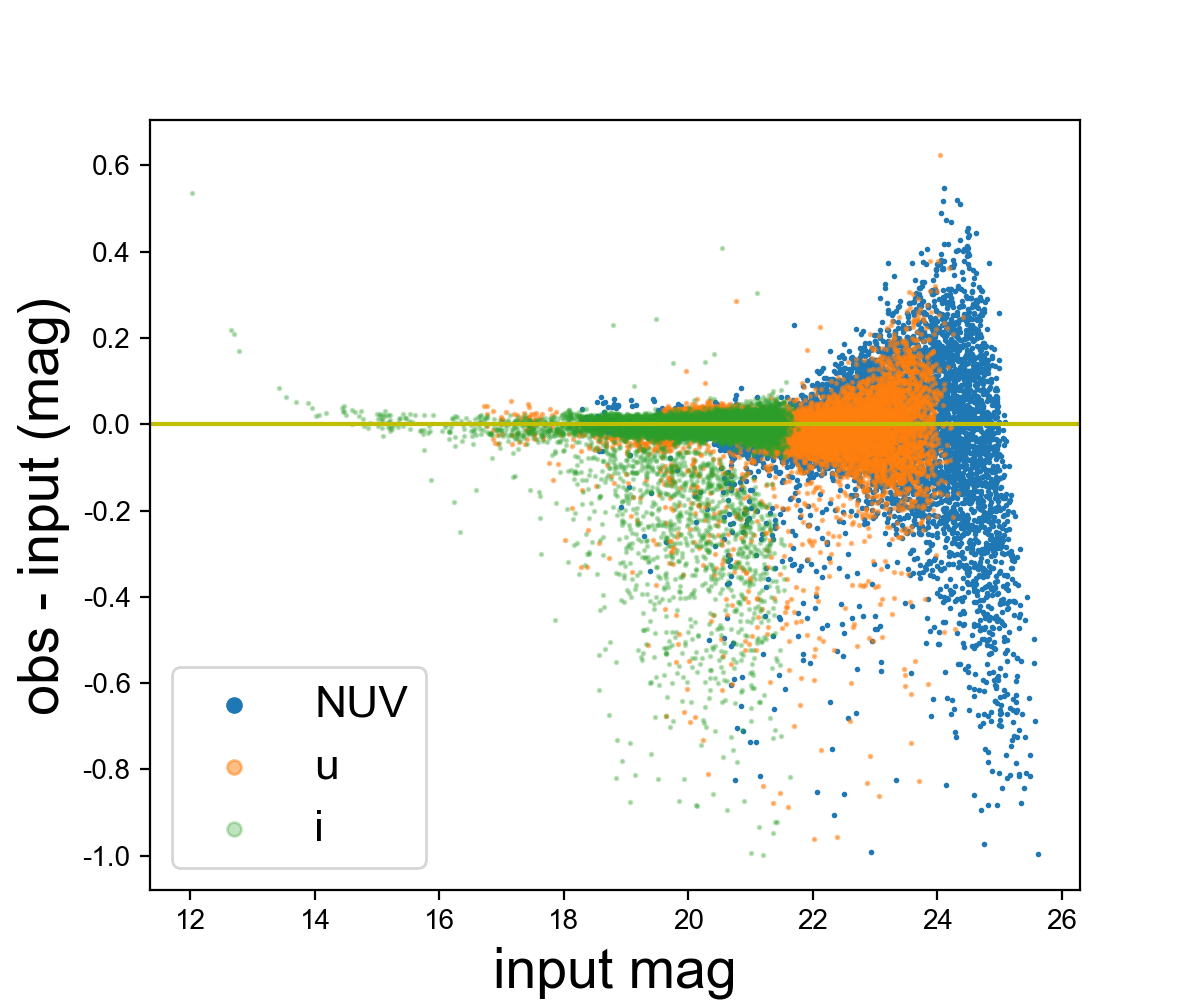}
      \caption{Photometric error curves for different filter band: blue for $NUV$-band($7\times150s$), orange for $u$-band($3\times50s +4\times150s$), and green for i-band($3\times10s +2\times50s+2\times150s$)}
         \label{fig7}
    \end{figure}

    \begin{figure*}[h!]
    \centering
    \includegraphics[width=\hsize]{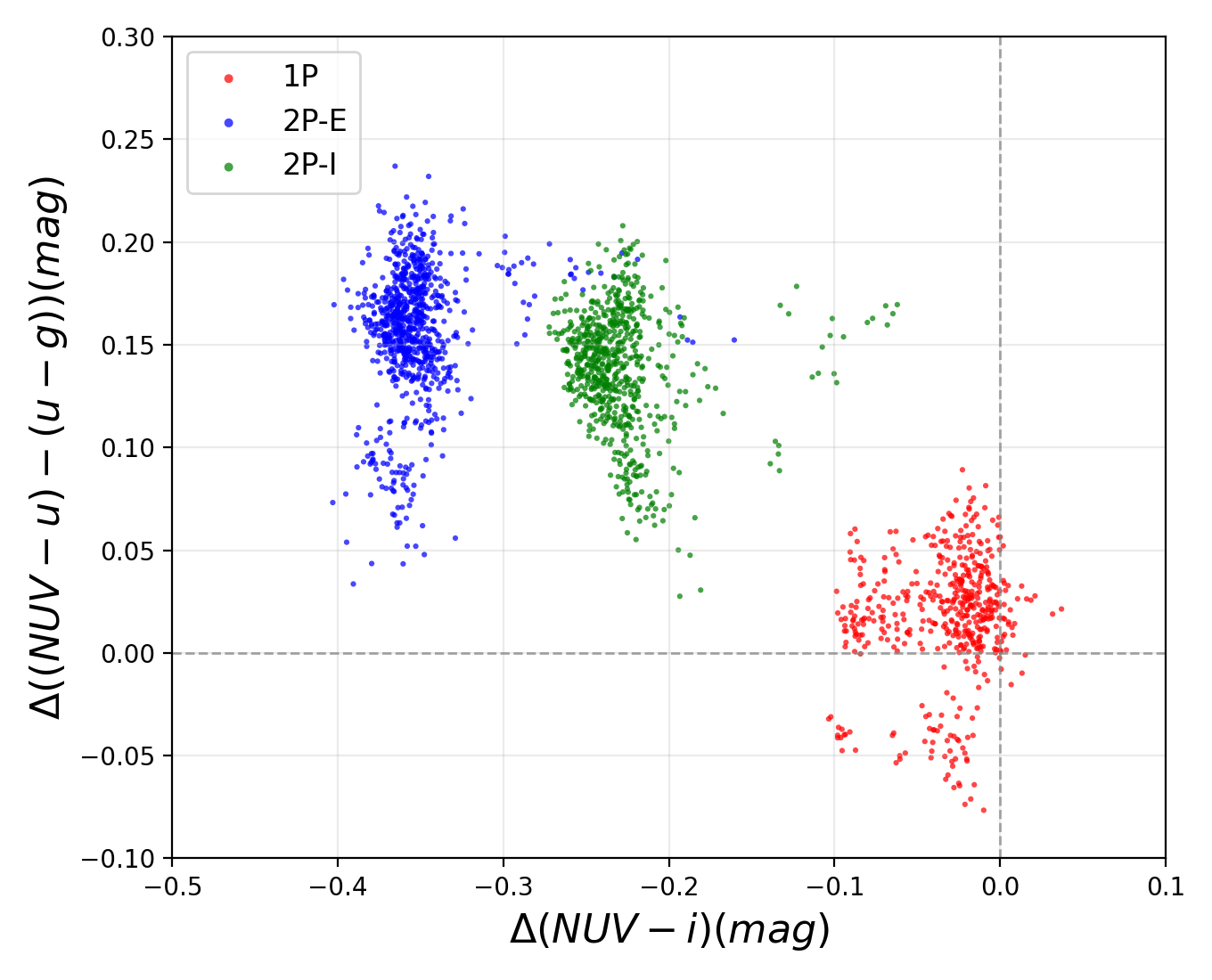}
      \caption{Chromosome maps of three populations shown in ~\ref{fig6}, using $\Delta$ $(NUV - u)-(u-g)$ as the vertical axis and $\Delta$ $(NUV-i)$ as the horizontal axis. Dashed lines indicate the red fiducial lines.}
         \label{fig8}
    \end{figure*}

   \begin{figure*}[h!]
        \centering
        \begin{minipage}{0.49\hsize}
            \centering
            \includegraphics[width=\linewidth]{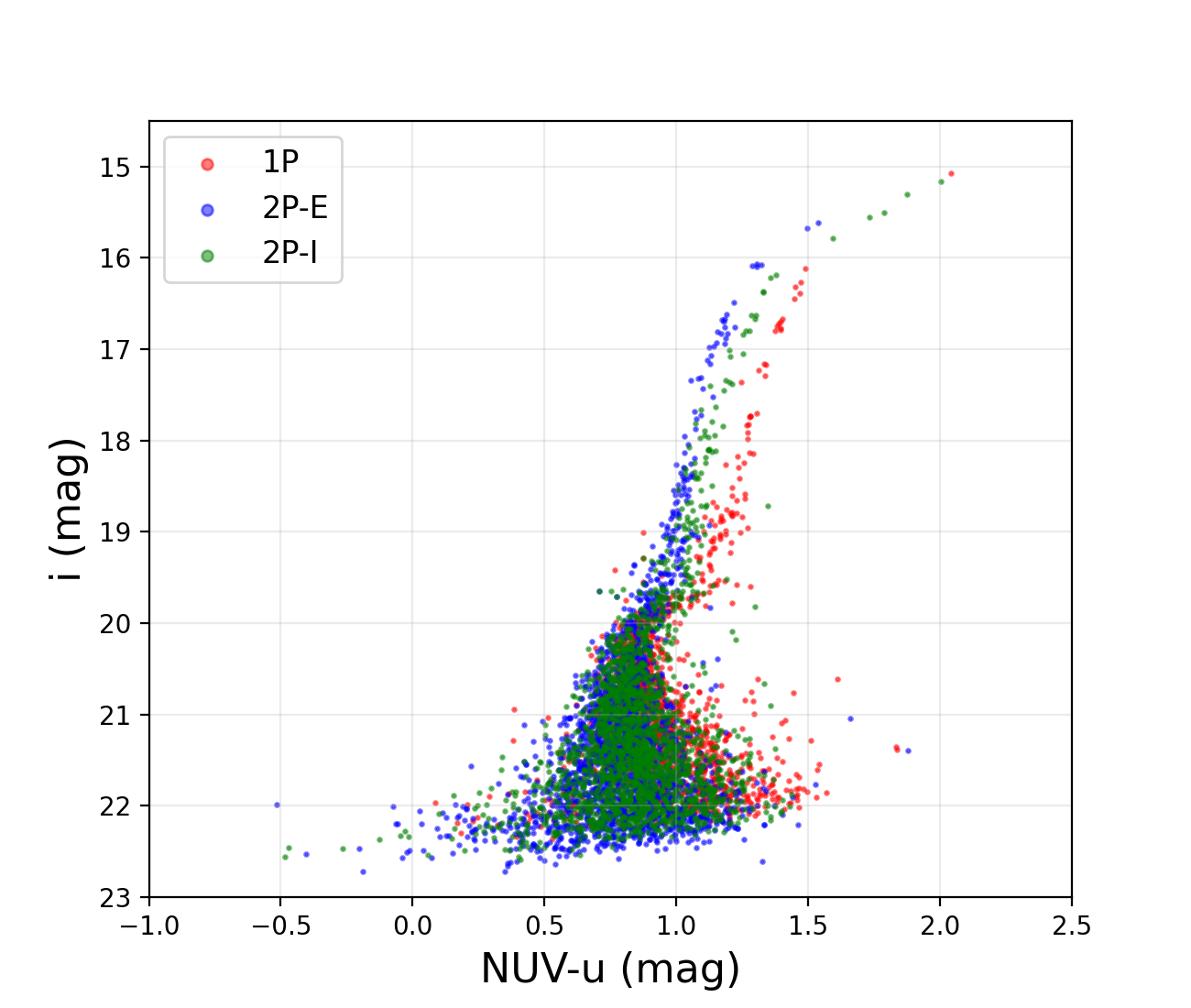}
        \end{minipage}
        \begin{minipage}{0.49\hsize}
            \centering
            \includegraphics[width=\linewidth]{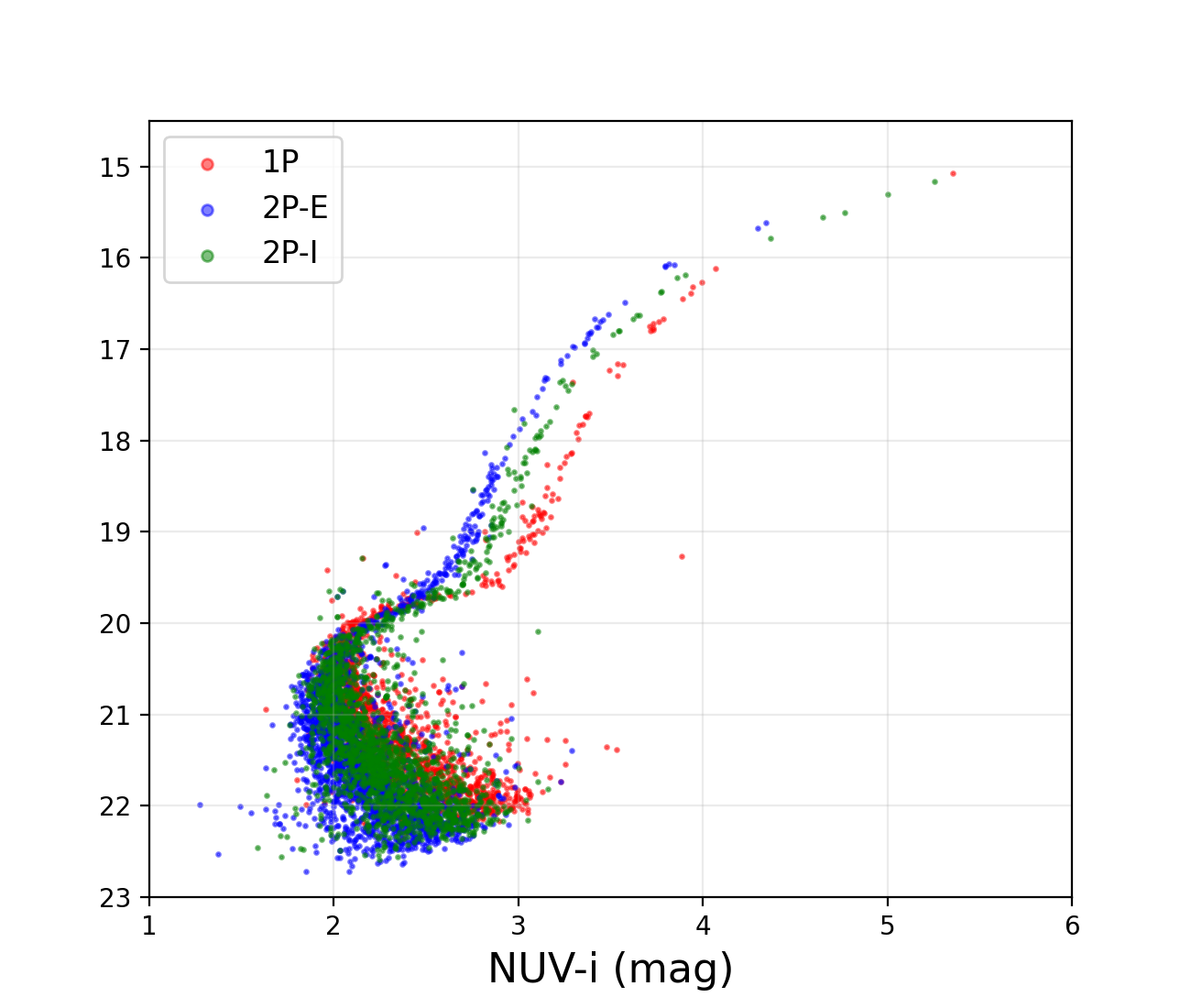}
        \end{minipage}
        \caption{\rev{CMDs for the same multiple stellar populations as in Fig.~\ref{fig6} (Dartmouth models), but located at a distance of 20~kpc. Only 10\% of the stars are shown for clarity. Left: $NUV-u$ vs $i$; Right: $NUV-i$ vs $i$. Note the fainter magnitude limit compared to Fig.~\ref{fig6} due to the increased distance.}}
        \ContinuedFloat 
        \label{cmd_dm_20}
    \end{figure*}

    \begin{figure*}[h!]
    \centering
    \includegraphics[width=\hsize]{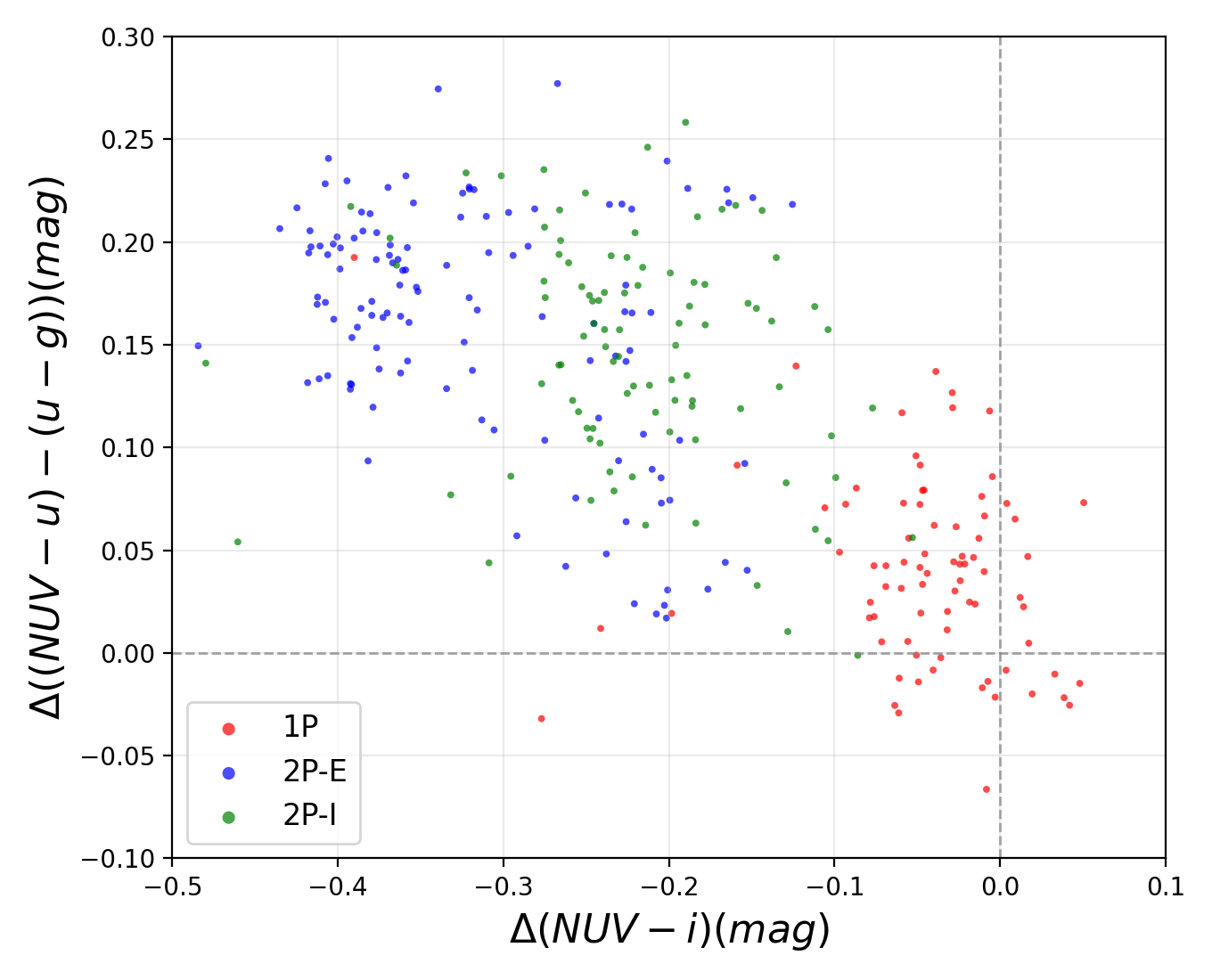}
      \caption{\rev{Chromosome maps of the three populations shown in Fig.~\ref{cmd_dm_20} at 20~kpc. The horizontal axis represents $\Delta(NUV-i)$ (sensitive to He abundance), and the vertical axis represents $\Delta[(NUV-u)-(u-g)]$ (sensitive to CNO abundances). Dashed lines indicate the red fiducial lines.}}
         \label{Chms_dm_20}
    \end{figure*}

   \begin{figure*}[h!]
        \centering
        \begin{minipage}{0.49\hsize}
            \centering
            \includegraphics[width=\linewidth]{ 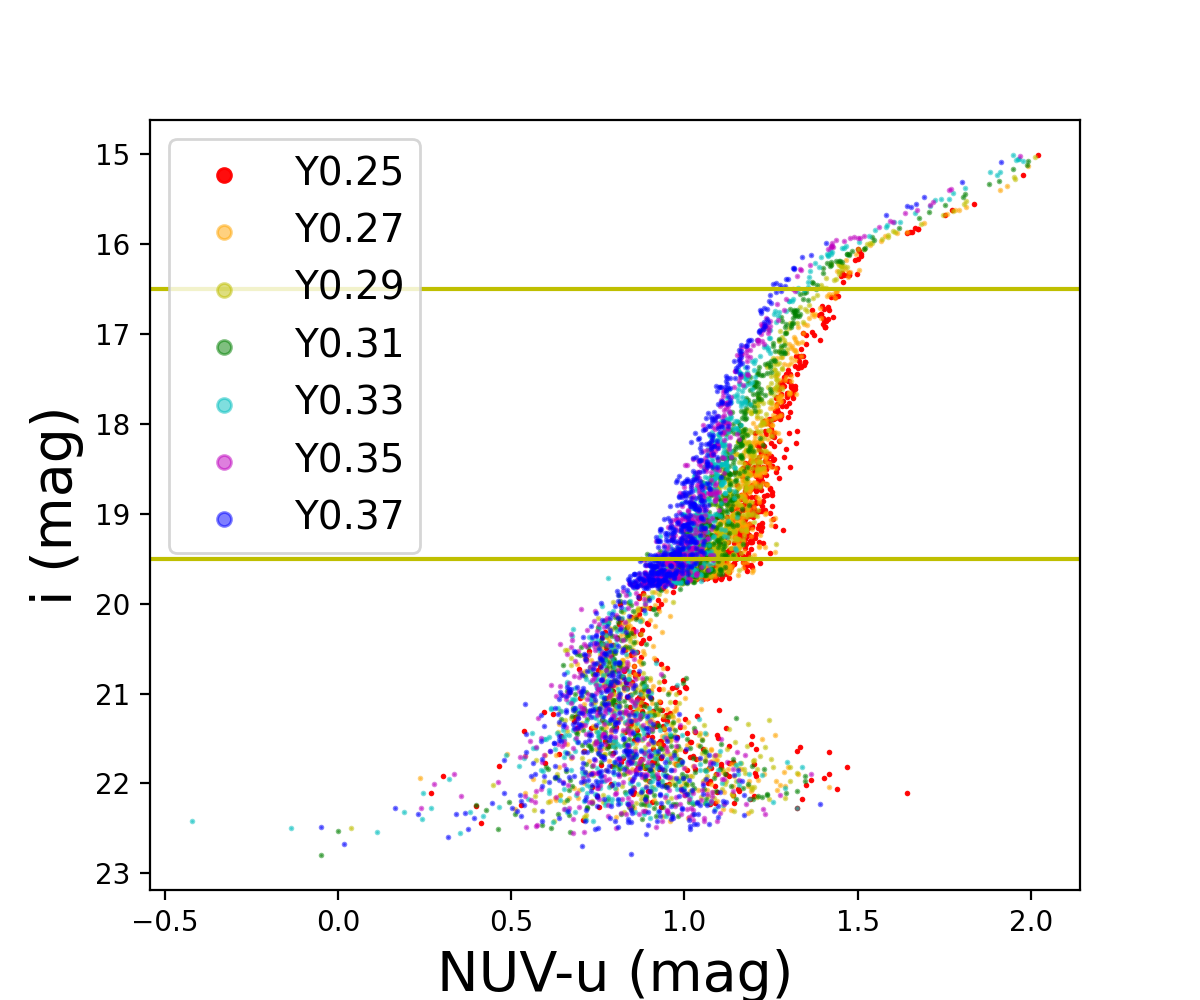}
        \end{minipage}
        \begin{minipage}{0.49\hsize}
            \centering
            \includegraphics[width=\linewidth]{ 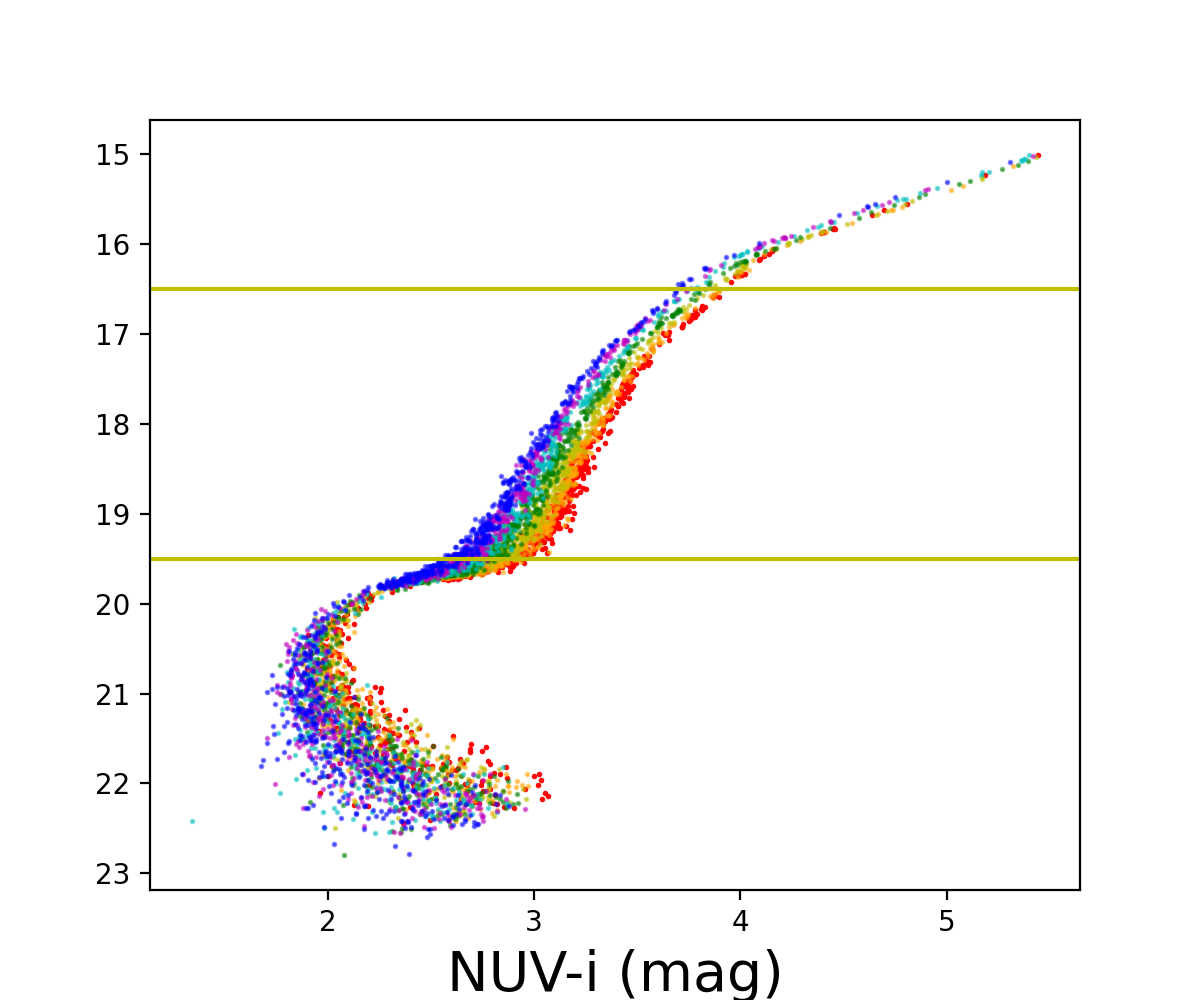}
        \end{minipage}
        \caption{\rev{CMDs for MPs with different chemical abundances defined in Table~\ref{tab:abundance_change} (PGPUC models) at 20~kpc. The seven populations span helium abundances from $Y=0.25$ (red) to $Y=0.37$ (blue). Left: $NUV-u$ versus $i$; Right: $NUV-i$ versus $i$.}}
        \ContinuedFloat 
        \label{fig9}
    \end{figure*}

    \begin{figure*}[h!]
    \centering
    \includegraphics[width=\hsize]{ 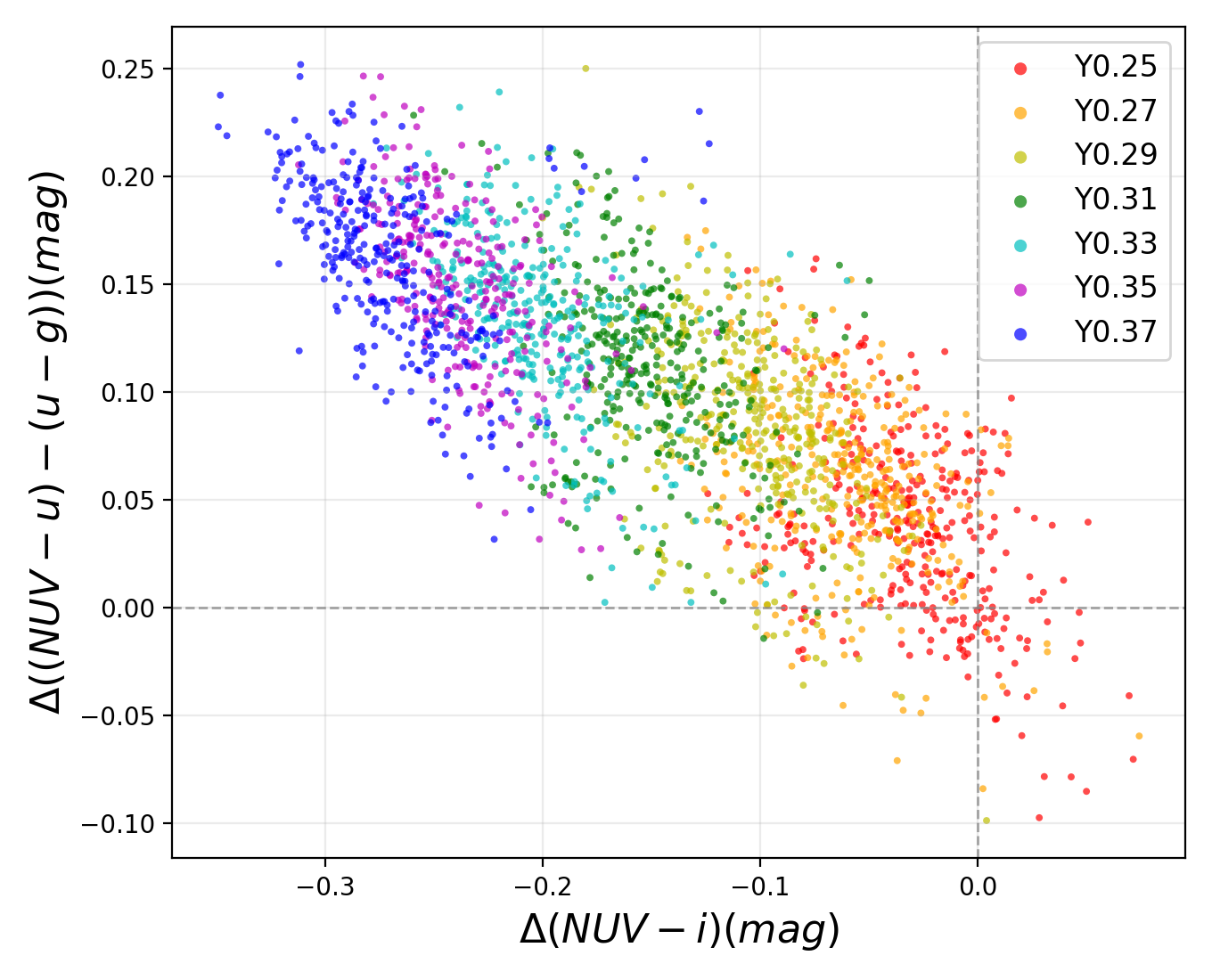}
      \caption{\rev{Chromosome maps of MPs defined in Table~\ref{tab:abundance_change} at 20~kpc, demonstrating the continuous sensitivity to helium abundance variations. Dashed lines show the red fiducial lines (corresponding to the $95^{\rm th}$ percentile of the $\Delta(NUV-i)$ and $\Delta[(NUV-u)-(u-g)]$ distributions for each bin in $i$). Populations with $\Delta Y \geq 0.06$~dex and $\delta[\rm N/Fe] \geq 0.64$~dex can be clearly resolved.}}
         \label{fig10}
    \end{figure*}

Then, this catalogue was put into the \texttt{CSST-msc-sim} simulation program(version~3.3.0, \cite{2026RAA....26b4001W})\footnote{\url{https://csst-tb.bao.ac.cn/code/csst-sims/csst_msc_sim}}. For the simulated PSF model, we employed a model that samples and interpolates the Huygens PSF after ray tracing, based on \cite{2025arXiv251106936B}. Dark current, bias, and readout noise were added to the simulation. Then we obtained simulated observation images (FITS files) for each filter band and a corresponding catalogue containing instrumental magnitudes, celestial coordinates, image pixel coordinates, and other relevant information (such as ID of filter). The exposure times were designed to balance depth and dynamic range. Given that the standard single-exposure duration for the CSST wide-field survey is fixed at $150$~s, we adopted this value as the maximum integration time for individual frames. The exposure times were set as follows: the $NUV$ band was exposed for $150$~s, repeated $7$ times; the $u$ band was exposed $3$ times for $50$~s and $4$ times for $150$~s; the $g$ and $i$ bands were each exposed $3$ times for $10$~s, $2$ times for $50$~s, and $2$ times for $150$~s, aiming to maximise the signal‑to-noise ratio (SNR) for faint stars while avoiding saturation of bright stars. The inclusion of short exposures pushes the saturation limit brighter than 14 magnitude, minimizing data loss for bright stars. It is worth noting that exposures shorter than 10 s were avoided, as the 3 s shutter movement time would introduce significant photometric non-uniformities (shutter effects). The exposure times adopted above represent a conservative estimate of the potential observing strategies. We acknowledge that, at least for calibration clusters, the actual exposure times provided by the CSST will significantly exceed the values assumed here. Photometry was performed on the obtained simulated images using the \texttt{kron} aperture mode of \texttt{Photutils} \citep{larry_bradley_2025_14889440}. Based on the multiband photometric magnitudes, we constructed a CMD of $NUV - u$ vs $i$ for 1P and 2P (2P-E \& 2P-I) stars, shown in Fig.~\ref{fig6}. We also constructed a CMD using $NUV-i$ vs $i$, shown in Fig.~\ref{fig_cmdd_n-i}, which enables a better distinction between stars with different helium contents. According to the isochrone theory, for stellar populations with different helium contents, the distinction between the $NUV-y$ and $NUV-z$ colours will be more effective than that of the $NUV-i$ colour. However, due to the lower transmission of $y$-band and $z$-band filters(as shown in Fig.~\ref{fig1}), we still used $NUV-i$. Photometric errors were determined from the differences between instrumental and photometric magnitudes, as presented in Fig.~\ref{fig7}. Photometric errors are primarily caused by photon (Poisson) noise and background noise, while blending effects due to closely spaced sources have been removed from the data. The zero-point offset was determined from the average magnitude of stars located 1–2 mag below the saturation limit, after two iterations of outlier rejection.

The photometric error curves derived from our mock observations indicate that, under the adopted exposure strategy, the detection limit in the $NUV$ and $u$ bands is reached at approximately 24\,mag; at this level, 95\% of the stars exhibit photometric errors of $\sim 0.2-0.25$\,mag (4--5$\sigma$). At an apparent magnitude of $m <\sim 20$\,mag, the signal from MPs begins to dominate, manifesting as a MS that is noticeably broader than the spread induced by photometric errors alone. For the RGB at magnitudes brighter than $i \sim 18$\,mag, the separation between the 1P and 2P populations is clearly distinguishable. Our exposure strategy for simulated star clusters results in saturation of stars with $g$-band magnitude of about $g \leq 15$ mag. Although this does not affect our physical analysis of the mock cluster (we only focus on the segment of the RGB with luminosity below the RGB Bump), it implies that for closer GCs, a shorter exposure strategy should be adopted to prevent overexposure of bright red giants.

We constructed ChMs of the MPs following the procedure outlined by \cite{2017MNRAS.464.3636M}. Specifically, we determined the blue and red fiducial lines corresponding to the 5th and 95th percentiles of the stellar distribution at each magnitude, which were then used to verticalize and normalise the colour indices. In this work, we used $\Delta_{(NUV-i)}$ to represent differences in helium abundance between different populations and $\Delta_{((NUV-u)-(u-g))}$ to represent differences in CNO abundances, as shown in Fig.\ref{fig8}. Since the chemical abundances are identical for stars within each population, the dispersion size of each population on the chromosome map can be regarded as the dispersion caused by photometric errors. It is evident that the separation between different stellar populations is significantly larger than that caused by errors.

\subsection{Extended Simulations at 20 kpc}

To assess the capability of CSST/SCam for more distant clusters, we extended our simulations to a distance of 20 kpc, which encompasses approximately 80\% of the known Galactic globular clusters \citep{1996AJ....112.1487H}. We first re-simulated the Dartmouth-based populations (1P, 2P-E, and 2P-I) at this distance. All other simulation parameters (PSF model, noise characteristics) remained identical to the 9.6~kpc simulations.

Figures~\ref{cmd_dm_20} and \ref{Chms_dm_20} present the resulting CMDs and chromosome maps for these 20~kpc distant populations, showing a randomly selected subsample of 10\% of the stars for clarity. Despite the fainter apparent magnitudes ($i <19.5$~mag for the red giants), the multiple sequences remain clearly distinguishable in the $NUV-u$ or $NUV-i$ colour. The chromosome map (Fig.~\ref{Chms_dm_20}) demonstrates that the three populations (1P, 2P-I, and 2P-E) remain separated.

Furthermore, to overcome the limitation of the Dartmouth models---which provide only discrete and relatively coarse steps in helium abundance ($Y = 0.25, 0.33, 0.40$)---we employed the \texttt{PGPUC} code \citep{2013A&A...553A..62V} to simulate a more continuous and realistic variation in both helium and CNO elements. We generated seven stellar populations with an evolutionary age of 12~Gyr and helium abundance $Y$ ranging from 0.25 to 0.37 in steps of 0.02, following the abundance patterns defined in Table~\ref{tab:abundance_change}. For each population, we selected 500 red giants and 500 dwarfs, resulting in a total of 7,000 simulated stars.The simulated exposures were set as follows: $NUV$ and $u$ bands each with 9 exposures of $150~s$; $g$ and $i$ bands each with 5 exposures of $10~s$ and 4 exposures of $50~s$. All other simulation settings remained identical to those applied to the MPs generated using the Dartmouth models.

After performing photometry on the simulated images, we constructed the CMD of MPs shown in Figure~\ref{fig9}. Excluding the saturated sources, we selected red giants between 16.5 to 19.5~mag (corresponding to red giants with their evolutionary stages before the RGB Bump) to construct the chromosome map displayed in Figure~\ref{fig10}. 

Figure~\ref{fig9} demonstrates that even at 20~kpc, the CSST/SCam can resolve multiple main sequences spanning the full range of helium abundances ($Y=0.25$--0.37). The corresponding chromosome map (Fig.~\ref{fig10}) reveals a continuous distribution of populations in the $\Delta(NUV-i)$ versus $\Delta[(NUV-u)-(u-g)]$ plane. Our analysis indicates that when $\Delta Y > 0.06$~dex and $\delta[\rm N/Fe] > 0.64$~dex, different stellar populations can be clearly distinguished by the chromosome map, confirming the sensitivity of CSST/SCam to moderate chemical abundance variations in distant clusters.

\section{Conclusions}
\label{sect:conclusion}

In this paper, we systematically investigate the capability of multi-band photometry from CSST/SCam to resolve MPs. Taking NGC 2808 as the prototype, we constructed a model star cluster with significant helium abundance dispersion ($\Delta Y=0.25 - 0.4$) and CNO variations ($\delta{[\rm N/Fe]}=1.0\ \rm dex$, $\delta{[\rm C/Fe]}=\delta{[\rm O/Fe]}=-0.5\ \rm dex$). Our results indicate that the $NUV-i$ colour index of CSST/SCam is the most sensitive to helium variation, while for CNO variation, the most sensitive colour index is $NUV-u$. 

By constructing mock observations based on virtual simulations, we analysed the distributions of MPs with different chemical abundance dispersions on the CMD and chromosome map under different exposure time strategies. Our findings demonstrate that the ``chromosome map'', constructed from CSST/SCam multi-band photometric data with $(NUV - u)-(u - g)$ as the pseudo-colour and $NUV - i$ as the colour index, enables the most efficient resolution of MPs characterised by distinct chemical abundance patterns. Under our implemented exposure strategy (with a total exposure time of $>$1000 s for the UV passbands and $>$300 s for optical $g$ and $i$ bands), the chromosome map is capable of discriminating between two distinct stellar populations with $\Delta Y\geq0.04$ dex and $\delta{[\rm N/Fe]}\geq0.5$ dex. The capability of CSST/SCam to resolve MPs can be extended to the apparent magnitude of $i\sim20$ mag under our exposure strategy. Provided that the chemical differences between different stellar populations maintain a moderate level at this point ($\Delta Y=0.08$ dex, $\delta{[\rm N/Fe]}>0.7$ dex, and $\delta{[\rm O/Fe]}\ (\rm [C/Fe]) <-0.3$ dex), CSST can detect a MS band significantly wider than that expected from photometric errors.

When extending the simulated distance to 20~kpc---which encompasses approximately 80\% of the Galactic globular clusters---the chromosome maps can still significantly resolve distinct stellar populations with $\Delta Y \geq 0.06$~dex and $\delta[\rm N/Fe] \geq 0.64$~dex. Furthermore, at 20~kpc, the CMD remains effective in distinguishing different stellar populations down to $i \sim 19.5$~mag, provided that the chemical differences are substantial ($\Delta Y=0.08$~dex, $\delta[\rm N/Fe]>0.7$~dex, and $\delta[\rm O/Fe]$ or $\delta[\rm C/Fe$] $<-0.3$~dex).

We highlight that a significant advantage of CSST over HST lies in its much larger FoV. This enables a substantial improvement in the efficiency of searching for MPs and an expansion of the cluster parameter space. As can be seen from Figure~\ref{fig5}, the FoV of CSST/SCam is nearly 600 times that of HST UVIS/WFC3 ($1.1^\circ\times1.1^\circ$ vs. $162''\times162''$). The large FoV of CSST allows us to study the properties of MPs in different environments. It enables people to obtain a complete distribution map of MPs ranging from the cluster core to the outskirts and the tidal structures, thereby placing important constraints on the dynamical properties of MPs. Our results thus confirm the potential of CSST to conduct large-scale, homogeneous surveys of MPs in Galactic and Local Group clusters, providing a solid foundation for future observational strategies after the telescope's launch.

\begin{acknowledgements}
We acknowledge support from the NSFC Key Program (grant No. 12233010).
\end{acknowledgements}

\appendix                  

\bibliographystyle{raa}
\bibliography{bibtex}

\end{document}